\documentclass[12pt]{article}

\pdfoutput=1 

\usepackage{amssymb}

\usepackage{amsmath}

\usepackage{graphicx}

\usepackage{booktabs}

\usepackage{float}

\usepackage{url}

\usepackage{appendix}

\usepackage{makeidx}

\usepackage{hyperref}

\title{Model-based fuzzy time series clustering of conditional higher moments}

\author{\normalsize Roy Cerqueti\thanks{Roy Cerqueti: roy.cerqueti@uniroma1.it} \\ \normalsize Department of Statistics and Social Sciences, Sapienza University of Rome, Italy \\
	\normalsize School of Business, London South Bank University, United Kingdom	\\ \normalsize Massimiliano Giacalone\thanks{Massimiliano Giacalone: massimiliano.giacalone@unina.it} \\ \normalsize Department of Economics and Statistics, University of Naples "Federico II" \\
	\normalsize Raffaele Mattera\thanks{Raffaele Mattera: raffaele.mattera@unina.it} \\ \normalsize Department of Economics and Statistics, University of Naples "Federico II" \\} 

\begin{document}

\date{}

\maketitle

\begin{abstract}
	This paper develops a new time series clustering procedure allowing for heteroskedasticity, non-normality and model's non-linearity. At this aim, we follow a fuzzy approach. Specifically, considering a Dynamic Conditional Score (DCS) model, we propose to cluster time series according to their estimated conditional moments via the Autocorrelation-based fuzzy $\mathcal{C}$-means (A-FCM) algorithm. The DCS parametric modelling is appealing because of its generality and computational feasibility. The usefulness of the proposed procedure is illustrated using an experiment with simulated data and several empirical applications with financial time series assuming both linear and nonlinear models' specification and under several assumptions about time series density function.
\end{abstract}


\section{Introduction}

Time-series data are of interest due to their ubiquity in various areas ranging from science and engineering to finance and economics. While each time series consists of many data points, it can also be seen as a single object. Clustering -- i.e., grouping objects with maximum similarity with other objects within the same group -- is a useful tool for exploratory analysis as it identifies common structures within data. Moreover, clustering procedures for time series have interesting applications in finance. For example, trough clustering methods, it is possible to build portfolios of similar stocks (e.g. \cite{iorio2018}). Clustering complex objects like time series is a difficult task because this kind of data has particular features like serial correlation and usually are both noisy and heteroskedastic with shifts \cite{aghabozorgi2015}. From a methodological point of view, time series clustering methods can be divided into three main classes \cite{liao2005}: observation-based, feature-based and model-based clustering. The first group classifies time series on the ground of the raw data (see, e.g. \cite{afridi2020, coppi2010, deluca2011, durso2000, durso2004, kovsmelj1990}). In the series with different lengths, the observation-based clustering methods are usually built upon the so-called Dynamic Time Warping (DTW). This technique allows finding an optimal alignment between two given sequences of different length. (e.g. \cite{caiado2009}, \cite{durso2018}). The methods in the second group cluster objects according to some of their features, like the autocorrelation function \cite{alonso2006,caiado2006, durso2009} or the parameters of the wavelet decomposition (e.g. \cite{durso2012,maharaj2010}). The models of the third group assume that the time series are generated from the same statistical model, and group objects according to the estimated parameters. In this context, we mention \cite{piccolo1990} that proposed a distance measure between ARMA processes. This study has been extended by several authors (e.g. \cite{corduas2008, otranto2008}). Considering heteroskedasticity, \cite{otranto2008} proposed a distance measure between GARCH processes. Then, he clustered time series according to their estimated conditional and unconditional variances. Paper \cite{otranto2010} extended \cite{otranto2008} for multivariate financial time series. Moreover, \cite{durso2016} proposed a new robust approach for clustering heteroskedastic time series by introducing the fuzzy logic in a GARCH-based approach. In this method, each time series could be in more than one cluster with a certain level of probability following a partitioning around medoids (PAM) procedure. 
\\
This paper presents a clustering procedure of model-based type. 
\\
Recent papers that follow a model-based approach (e.g. \cite{otranto2008, otranto2010, durso2016}) propose to cluster time series according to conditional variance estimates. Time-varying variance is a crucial aspect of heteroskedastic time series, but it is not the only one we should look at to cluster dynamic objects. Indeed, it is reasonable to assume that the entire conditional distribution is time-varying in a time series framework. Some authors (e.g. \cite{deluca2011}) showed the usefulness of considering higher moments for clustering. In particular, \cite{deluca2011} developed a clustering procedure to group time series with similar tails.
\\
There are several statistical models able to capture the dynamic behaviour of conditional higher moments. Examples are \cite{harvey1999} and \cite{leon2005}. Recently, \cite{creal2013, harvey2013} developed the Dynamic Conditional Score (DCS also called Generalized Autoregressive Score), a very general statistical model that considers the score function of the predictive model density as the driving mechanism for time‐varying parameters. A wide class of GARCH-type processes are special cases of the DCS model (see, e.g. \cite{creal2013,ayala2017}). Since the DCS model is based on the score, it exploits the data's complete density structure rather than just a few moments. 
\\
Similarly to the proposal of \cite{otranto2008, otranto2010, durso2016}, which classifies time series according to the estimated conditional variances, it is possible to cluster time series according to the estimated conditional moments obtained via a Dynamic Conditional Score.
\\
One of the crucial points is to determine a similarity measure for the estimated conditional higher moments. Note that the estimated conditional moments are essentially time series themselves. Various distances have been proposed for clustering evolutive objects \cite{liao2005, aghabozorgi2015, diaz2010}. 
\\
We take inspiration from the works carried out by \cite{durso2009, durso2016} since we adopt a fuzzy logic within a model-based clustering procedure. In particular, we use the Autocorrelation-based fuzzy $\mathcal{C}$-means algorithm (the A-FCM of \cite{durso2009}) to cluster time series with similar conditional moments. Such conditional moments are estimated through a DCS procedure.
\\
In general, the fuzzy approach for clustering is very appealing for time series data, as shown by \cite{durso2016} that demonstrated its validity in model-based procedures.  The time series's dynamic behaviour plays an essential role in the clustering process, as argued by several authors (e.g. \cite{maharaj2010, maharaj2011, durso2018, maharaj2019}). The reasons for that are different. The fuzzy approach allows one time series to be allocated to two or more clusters with a membership degree that represents the uncertainty related to each time series assignment to each cluster. The membership degree implicitly indicates the presence of a second-best cluster almost as good as the first best. Traditional clustering methods are not able to highlight such conclusions.
Moreover, in the real world, identifying a clear boundary between clusters is not easy, so a fuzzy approach appears more attractive than a deterministic one. The literature also highlights that fuzzy clustering procedures are computationally more efficient \cite{durso2016}. In the end, the fuzzy logic has been proved very useful for clustering time series with particular properties \cite{cont2001}, like the financial one (\cite{maharaj2010b, durso2013, huynh2013}).
\\
Furthermore, thanks to the generality of the underlying statistical model, we are able to consider both linear and nonlinear model specifications and to cluster also non-Gaussian time series. Importantly, our approach considers time-varying higher moments and does not restrict only to the conditional variance.
\\
The present paper is structured as follows. In the next section, the statistical model underlying our model-based procedure is presented. Then, in Section \ref{sec3}, we discuss the fuzzy clustering algorithm in detail. To this aim, we first specify the employed distance measure and then the procedure used for estimating and clustering the conditional moments. The proposed method's usefulness is illustrated using several empirical applications with financial time series assuming both linear and nonlinear models' specification and under several assumptions about time series density function. In particular, Section \ref{sec4} presents the used dataset and outlines the methodologies, while Section \ref{sec5} collects and discusses the obtained empirical results. The figures related to the estimated conditional moments and their autocorrelation functions are postponed in a devoted Appendix for easy reading.

\section{Dynamic Conditional Score model}

\subsection{Basic statistical model}

Let us consider a time series $(y_t:t=1, \dots, T)$ with $f_t$ the time‐varying parameter vector observed at time $t$. In our context, the available information set at time $t$ is $\mathcal{F}_t$ and -- as we will see below -- it is given by a collection of some previous realizations of the $f$'s and the $y$'s. Then, define $\theta$ a vector of static parameters. In the Dynamic Conditional Score model we assume that $y_t$ is generated by the observation density:

\begin{equation}
 \label{eq:1}
    y_t \sim p(y_{t} | f_t, \mathcal{F}_t; \theta),
\end{equation}
where the dependence on the parameter $\theta$ -- which is conveniently let be explicit in (\ref{eq:1}) -- is due to the dependence of $f_t$ on $\theta$. 

Indeed, given two integers $ 0 \leq n , m  \leq  T-1$, we formalize the $DCS(n,m)$ model for the $t$-th realization $f_t$ of the time-varying parameter vector as:

\begin{equation}
 \label{eq:2}
    f_t = \omega + \sum_{i=1}^{n} \mathbf{A}_i s_{t-i} + \sum_{j=1}^{m} \mathbf{B}_j f_{t-j}
\end{equation}
\\
where $\omega$ is a real vector and the $\mathbf{A}$'s and the $\mathbf{B}$'s are real matrices with an appropriate dimension -- all the scalar parameters in $\omega, \mathbf{A}_1, \dots, \mathbf{A}_n, \mathbf{B}_1, \dots, \mathbf{B}_m$ are collected in the vector $\theta$ introduced in (\ref{eq:1}); $s_t$ is a type of score of the conditional distribution defined in (\ref{eq:1}), and it is a function of the data and the parameters, so that $s_t=s_t(y_t, f_t, \mathcal{F}_t; \theta)$. More specifically, this class of models suppose that the evolution of the time-varying parameter vector $f_t$ is driven by a vector proportional to the score of density (\ref{eq:1}) together with an autoregressive component. Indeed, $s_t$ is defined as:

\begin{equation}
 \label{eq:3}
    s_t = S_t\cdot \nabla_t 
\end{equation}
\\
where $S_t = S_t(f_t, \mathcal{F}_t; \theta)$ is a positive definite scaling matrix known at time $t$ and $\nabla_t (y_t, f_t,\mathcal{F}_t; \theta)$ is the score of $y_t$ evaluated with respect to $f_t$, i.e.:

\begin{equation}
 \label{eq:4}
    \nabla_t = \frac{\partial \log p(y_{t} | f_t, \mathcal{F}_t; \theta)}{\partial f_t}
\end{equation}

Since the score depends on the complete density and not only on some moments of $y_t$, the $DCS(n,m)$ model uses the full density structure for updating $f_t$. We have to highlight that we could get different $DCS(n,m)$ specifications depending by the choice about scaling $S_t$ we make. A very appealing feature of $DCS(n,m)$ model is that the vector of parameters $\theta$ could be estimated by maximum likelihood as the solution of the following problem (see e.g. \cite{creal2013}):

\begin{equation}
 \label{eq:5}
    \hat{\theta} = \arg \max_{\theta} \sum_{t=1}^{T} \ell_t
\end{equation}
\\
where $\ell_t = \log p(y_{t} | f_t, \mathcal{F}_t; \theta)$ for a realization of $y_t$.

\subsection{Some special cases}

As we said, the Dynamic Conditional Score model is very general since we could get different model specifications considering different assumptions about the scaling (\ref{eq:3}), the observation density (\ref{eq:1}) and $f_t$. A common approach, proposed by \cite{creal2013}, is to scale the score in order to account for its variance. Particularly, the authors proposed to scale using a matrix $S_t$ equal to the inverse of the information matrix of $f_t$ to a power $\gamma  \geq 0$:

\begin{equation}
 \label{eq:6}
    S_t = \mathcal{I}^{-\gamma}
\end{equation}
\\
with $\mathcal{I}$:

\begin{equation}
 \label{eq:7}
    \mathcal{I} = E \left[ \nabla_t \nabla_t' \right]
\end{equation}
\\
where $E$ denotes the expectation at time $t-1$ and the conditional score $\nabla_t$ is defined as (\ref{eq:4}). The parameter $\gamma$ is fixed by the user and usually takes value in the set $\{0,\frac{1}{2}, 1\}$. When $\gamma=0$, then $S_t$ is the identity matrix -- i.e., $S_t = \mathbf{I}$ -- and there is no scaling. Differently, if $\gamma=1$, then the conditional score $\nabla_t$ is premultiplied by the inverse of (\ref{eq:7}) to obtain (\ref{eq:3}). Finally, $\gamma=\frac{1}{2}$ leads to $\nabla_t$ scaled to its square-root to obtain (\ref{eq:3}).
\\
We notice that setting $\gamma=0$ one has that the $DCS(n,m)$ model contains the Autoregressive Conditional Multinomial (ACM) model of \cite{russell2005} and the GARMA models of \cite{benjamin2003} as particular cases.
\\
If we set $f_t = \sigma^2_t$, by scaling the score with (\ref{eq:6}) and setting $\gamma=1$, then the $DSC(n,m)$ model results in the well-known GARCH(n,m) model of \cite{bollerslev1986}. Particularly, assuming a Gaussian density in (\ref{eq:1}), the $DCS(1,1)$ becomes a Gaussian GARCH(1,1) (see \cite{creal2013}).
\\
If we suppose $f_t = log(\sigma^2_t)$ from a $DCS(1,1)$ we get Exponential GARCH (E-GARCH) of \cite{nelson1991}.
\\
Differently, if we suppose t-student density with $v$ degrees of freedom for $y_t$ we get the Beta-t-EGARCH of \cite{harvey2008} and \cite{harvey2014}.
Interestingly, the Beta-t-EGARCH is different from the t-GARCH(1,1) model of \cite{bollerslev1987}. Indeed, differently \cite{bollerslev1987}, the $t-DCS(1,1)$ does not results in a substantial increase in the variance after a large absolute change in $y_t$ . 
\\
Another interesting example to mention is when we have $y_t \sim G.E.D.(\mu_t, \phi_t, v_t)$. Indeed, assuming in this case $f_t = \sigma^2_t$ we obtain the GED-GARCH(m,n) process \cite{wisniewska2017}, while letting all parameters $f_t = (\mu_t, \phi_t, v_t)$ to be time vary-varying we get the $GED-DCS(n,m)$ model \cite{ayala2017}.
\\
A very important feature of this family of distributions is that they include also other common distributions, for different values of shape parameter $v$. This assumption has been revealed as very useful for modeling financial time series (e.g. \cite{cerqueti2020},\cite{cerqueti2019},  \cite{nelson1991}, \cite{theodossiou2015a}).
\\
As we can see, the $DCS(n,m)$ could be in principle nonlinear and able to model conditional heteroskedastic time series as well as time series with conditional time-varying higher moments. 

\section{Fuzzy clustering of DCS-based conditional moments}
\label{sec3}

We adopt a fuzzy approach in order to account for the uncertainty in time series clustering. 
The time clustering criterion is grounded on the conditional higher moments. 
\\
The outline of the procedure is as follows. We first estimate $R$ conditional moments according to the Dynamic Conditional Score model presented in the previous section, hence adopting a model-based procedure. Then, we cluster time series according to the features of the estimated conditional moments. In details, time series with similar $r$-th conditional moment -- where similarity is measured through a predefined concept of distance, see the next subsection -- are placed in the same cluster, for each $r=1, \dots, R$. Hence, by estimating $R$ moments, we can define $R$ levels of clustering (\cite{otranto2008}). Within each level $r=1, \dots, R$, we cluster time series according to their conditional moment over time path similarity. Statistical reasoning suggests that once two series belong to the same cluster for all the $R$ conditional moments, then they share similar conditional distribution. This remark is also in the light of authoritative studies (see, e.g. \cite{otranto2008, otranto2010}). 

The adopted distance measure is of Euclidean type. Specifically, we introduce a similarity criterion based on the Euclidean distance between a sufficiently large set of estimated autocorrelations.

Let define $\hat{\rho}^{(r)}_{l,k}$ the estimated autocorrelation at lag $l$ of the conditional $r$-th moment for a given $k$-th time series. Estimated autocorrelation $\hat{\rho}^{(r)}_{l,k}$ could be obtained with the usual estimator:

\begin{equation}
\label{eq:10}
\hat{\rho}^{(r)}_{l,k} = \frac{\sum_{t=l+1}^{T} (y^{(r)}_{k,t} - \bar{y}^{(r)}) (y^{(r)}_{k,t-l} - \bar{y}^{(r)})}{\sum_{t=1}^{T} (y^{(r)}_{k,t} - \bar{y}^{(r)})^2}
\end{equation}
\\
where $\bar{y}^{(r)}$ is the process' mean. A simple distance between two time series $X$ and $Y$ of $r$-th conditional moment that is based on estimated conditional moments' autocorrelation can be defined as:

\begin{equation}
 \label{eq:11}
    d^{(r)}_{ACF}(X, Y) =  
    \sum_{l=1}^{L} \left(\hat{\rho}^{(r)}_{l,X}-\hat{\rho}^{(r)}_{l,Y}\right)^2, 
\end{equation}
\\
that is equal to the squared Euclidean distance between the estimated autocorrelation functions. In using the distance defined in (\ref{eq:11}), we account for the autocorrelation structure of the processes, hence being in line with the fuzzy clustering approach proposed in \cite{durso2009}.


The procedure we propose is based on the fuzzy $\mathcal{C}$-means (FCM) algorithm \cite{krishnapuram2001, bezdek2013}. In particular, we adopt a fuzzy clustering algorithm for time series with similar autocorrelation structure of conditional moments using the autocorrelation-based fuzzy $\mathcal{C}$-means (A-FCM) algorithm of \cite{durso2009}. 
\\
The proposed clustering model could be formalized as follows:

\begin{equation}
\label{eq:model}
    \min: \sum_{k=1}^{K} \sum_{c=1}^{C} u^m_{k,c} \sum_{l=1}^{L} \left(\hat{\rho}^{(r)}_{l,k}-\hat{\rho}^{(r)}_{l,c}\right)^2
\end{equation}
\\
under the constraints:

\begin{equation*}
    \sum_{c=1}^{C}u_{k,c}=1,u_{u,c} \geq 0 \quad \text{and} \quad -1 \leq \hat{\rho}^{(r)}_{l,c} \leq 1
\end{equation*}
\\
where $u_{k,c}$ denotes the membership degree of the $r$-th conditional moments of the $k$-th time series to the $c$-th cluster, the parameter $m > 1$ controls for the fuzziness of the partition, $\hat{\rho}^{(r)}_{l,k}$ is the k-th unconditional moments' autocorrelation at $l$-th lag and $\hat{\rho}^{(r)}_{l,c}$ represents the $c$-th centroid autocorrelation. 
\\
Following \cite{durso2009}, the optimal solutions of the model (\ref{eq:model}) are equal to:

\begin{equation*}
    u_{k,c} = \frac{1}{\sum_{c'=1}^{C}\left( \frac{\sum_{l=1}^{L} (\hat{\rho}^{(r)}_{l,k}-\hat{\rho}^{(r)}_{l,c})^2 }{\sum_{l=1}^{L}} (\hat{\rho}^{(r)}_{l,k}-\hat{\rho}^{(r)}_{l,c'})^2\right)^{1/(m-1)}}
\end{equation*}
\\
and:

\begin{equation*}
    \hat{\rho}^{(r)}_{l,c} = \frac{\sum_{k=1}^{K}u_{k,c}^m \hat{\rho}^{(r)}_{l,k}}{\sum_{k=1}^{K} u_{k,c}^m}
\end{equation*}
\\
To solve the constrained optimization problem in (\ref{eq:model}), let us consider the following Lagrangian function: 

\begin{equation*}
 \mathcal{L}(u_{k,c}, \lambda) = \sum_{k=1}^{K} \sum_{c=1}^{C} u^m_{k,c} \sum_{l=1}^{L} (\hat{\rho}^{(r)}_{l,k}-\hat{\rho}^{(r)}_{l,c})^2 - \lambda \left(\sum_{c=1}^{C} u_{k,c} - 1 \right)
\end{equation*}
\\
Then it follows that:

\begin{align*}
    \frac{\partial \mathcal{L}(u_{k,c},\lambda)}{\partial u_{k,c}}=0\Longleftrightarrow m u_{k,c}^{m-1} \sum_{l=1}^{L} (\hat{\rho}^{(r)}_{l,k}-\hat{\rho}^{(r)}_{l,c})^2 - \lambda = 0\\
     \frac{\partial \mathcal{L}(u_{k,c},\lambda)}{\partial \lambda}=0\Longleftrightarrow \sum_{c=1}^{C}u_{k,c}-1=0
\end{align*}
\\
By substitutions of the two equations above we get the solution for the membership degree $u_{k,c}$. Similarly it could be shown that by taking the derivative of the Lagrangian function above with respect to $\hat{\rho}^{(r)}_{l,c}$ we get the solution showed before.
\\
Moreover, we have to highlight that the solution of $\hat{\rho}^{(r)}_{l,c}$ satisfies for the internality property the autocorrelation constraints $-1 \leq \hat{\rho}^{(r)}_{l,c} \leq 1$.
\\
Summarising, the procedure we propose involves the following steps:

\begin{enumerate}
    \item For each $k$-th time series $(y_{k,t}:t \geq 0)$, we estimate the DCS model with its parameters.
    \item Then, we provide in sample predictions on the ground of step 1. in order to obtain the $r$-th conditional moment of $y_{k,t}$ -- hence obtaining the time series $(\Tilde{y}^r_{k,t}: t \geq 0)$ -- for each $r=1, 2, \dots, R$.
    \item For any couple of conditional moment time series $(\Tilde{y}^r_{k_1,t}:t \geq 0)$ and $(\Tilde{y}^r_{k_2,t}:t \geq 0)$, we store the dissimilarities based on the distance defined in (\ref{eq:11})\footnote{In the Section 6 with simulated data we study the effect of different lags' selection. We suggest to select a lag length of $L=50$ in computing the dissimilarity measure. Indeed, also the simulations of \cite{diaz2010} show that $L=50$ provides the lowest miss-classification rate for autocorrelation-based distances, especially for large time series.}.
    \item We use the autocorrelation-based fuzzy $\mathcal{C}$-means (A-FCM) algorithm of \cite{durso2009} in order to generate clusters, for each $r=1,2, \dots, R$.
\end{enumerate}

At the end of the procedure, we could define $r$-levels of clusters depending on how many conditional moments we estimate. Then, we could be able to understand also which time series are located in the same cluster for more than one conditional moment. In doing so, we obtain a clustering of the considered time series.

\section{Empirical experiments: data and methods}
\label{sec4}

This section outlines the main ingredients of the proposed empirical application.

\subsection{Data}

\begin{figure}[H]
    \centering
    \includegraphics[width=\textwidth]{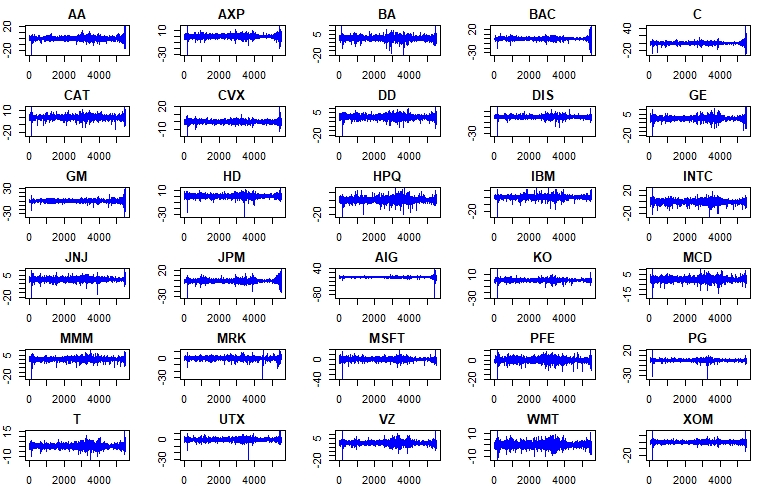}
    \caption{Stock returns' time series}
    \label{fig:ts1}
\end{figure}

We consider the dataset of \cite{ardia2019}, consisting in daily log-returns of the Dow Jones 30 constituents from 3/03/1987 to 3/02/2009 for a total of $T=5,521$ observations for each $k$-th series\footnote{Data and the R code can be retrieved at the following link: \url{https://www.sites.google.com/view/raffaele-mattera/} in the \emph{Research} Section.}. The constituents are labelled according to the standard nomenclature. Time series plots are shown in Fig. \ref{fig:ts1}.

\begin{figure}[H]
    \centering
    \includegraphics[width=\textwidth]{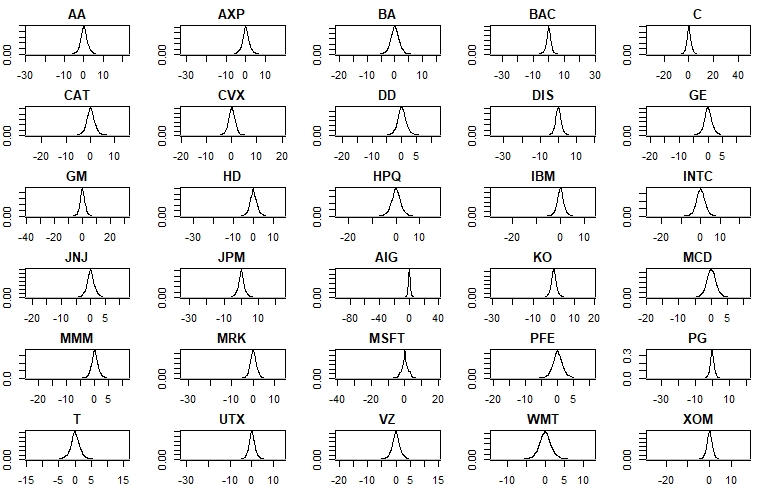}
    \caption{Stock returns' empirical densities}
    \label{fig:ts2}
\end{figure}

All the considered stock returns are stationary. Indeed, the absence of a deterministic trend is clear from Fig. \ref{fig:ts1}. Moreover, an unreported analysis shows also the absence of a unit root, hence getting additional evidence of the stationarity. We notice that the Dow Jones 30 constituents do not share the same variability since some of them have more a volatile pattern. Then, most of them are far from the origin on the y-axis. However, we could not say anything more about data distribution. A common approach is to assume normality. However, real data are far to be normally distributed, especially financial returns. Indeed, as we can see from stock returns' empirical densities in Fig. \ref{fig:ts2}, data are not normally distributed. This consideration is valid for almost all of them.

Specifically, they show heavy tails and different degrees of skewness. As further evidence, in the following Tab. \ref{tab:t5} are reported the results of Jarque-Bera (JB) normality test \cite{jarque1987} for each $k$-th time series. Considering the empirical kurtosis and skewness, under the null hypothesis of the test, we have that observed data are generated by a normal distribution. According to \cite{thadewald2007}, both \cite{jarque1987} and its adjusted version developed by \cite{urzua1996} are on the whole, the most powerful normality tests.

\begin{table}[H]
\centering
\begin{tabular}{rrr}
  \hline
 & JB statistic & Adj. JB statistic \\ 
  \hline
AA & 30910.08$^{***}$ & 30999.78$^{***}$ \\ 
  AXP & 28755.23$^{***}$ & 28838.50$^{***}$ \\ 
  BA & 13456.26$^{***}$ & 13496.51$^{***}$ \\ 
  BAC & 184045.41$^{***}$ & 184558.98$^{***}$ \\ 
  C & 244742.40$^{***}$ & 245424.44$^{***}$ \\ 
  CAT & 15772.73$^{***}$ & 15819.45$^{***}$ \\ 
  CVX & 31776.80$^{***}$ & 31869.05$^{***}$ \\ 
  DD & 10309.66$^{***}$ & 10340.85$^{***}$ \\ 
  DIS & 99929.77$^{***}$ & 100211.05$^{***}$ \\ 
  GE & 17217.36$^{***}$ & 17268.28$^{***}$ \\ 
  GM & 131518.47$^{***}$ & 131887.79$^{***}$ \\ 
  HD & 53299.40$^{***}$ & 53450.78$^{***}$ \\ 
  HPQ & 9358.56$^{***}$ & 9387.15$^{***}$ \\ 
  IBM & 40738.90$^{***}$ & 40856.05$^{***}$ \\ 
  INTC & 9354.40$^{***}$ & 9382.69$^{***}$ \\ 
  JNJ & 28788.97$^{***}$ & 28872.46$^{***}$ \\ 
  JPM & 47471.39$^{***}$ & 47607.51$^{***}$ \\ 
  AIG & 14974628.73$^{***}$ & 15015384.12$^{***}$ \\ 
  KO & 108544.37$^{***}$ & 108849.92$^{***}$ \\ 
  MCD & 8583.08$^{***}$ & 8609.35$^{***}$\\ 
  MMM & 35831.43$^{***}$ & 35934.46$^{***}$ \\ 
  MRK & 94355.75$^{***}$ & 94620.31$^{***}$ \\ 
  MSFT & 60033.02$^{***}$ & 60203.46$^{***}$ \\ 
  PFE & 6114.66$^{***}$ & 6133.65$^{***}$ \\ 
  PG & 1136387.62$^{***}$ & 1139501.35$^{***}$ \\ 
  T & 5349.30$^{***}$ & 5366.26$^{***}$ \\ 
  UTX & 87436.73$^{***}$ & 87682.86$^{***}$ \\ 
  VZ & 13369.83$^{***}$ & 13410.02$^{***}$ \\ 
  WMT & 2545.61$^{***}$ & 2554.20$^{***}$ \\ 
  XOM & 119180.17$^{***}$ & 119515.22$^{***}$ \\ 
   \hline
\end{tabular}
\caption{Normality tests. *** mean significance at 99\% of confidence level.}
\label{tab:t5}
\end{table}

For all stock returns, we reject the null. If we suppose that higher moments of the data distribution could be time-varying, these findings justify different modelling approaches that are discussed in the following subsection.

\subsection{Methodology; employed classes of DCS models}

First, we consider a linear and Gaussian model specification. This kind of assumption is in line with other classical papers of model-based financial time series clustering (e.g. \cite{otranto2008, durso2016}). Then, we relax such a condition by considering the t-Dynamic Conditional Score  model -- also called t-Beta-EGARCH -- that is both non-Gaussian and nonlinear, and the skew-t-Dynamic Conditional Score (skew-t-DCS), that allows also for time-varying skewness in asset returns. In the first case, we have two time-varying moments; in the second one, we have three conditional moments (location, scale and shape). For comparison purposes, in the third and last experiment, we estimate four conditional moments (location, scale, skewness and shape).

\subsubsection{Gaussian Dynamic Conditional Score}
\label{guassiandcs}
In the first application, we specify a Gaussian-DCS(1,1) model for all time series. In particular, we suppose $y_t \sim \mathcal{N}(\mu_t, \sigma^2_t)$ where:

\begin{equation*}
    p(y_t| f_t, \mathcal{F}_t; \theta) = \frac{1}{{\sigma_t \sqrt {2\pi } }}e^{{{ - \left( {y_t - \mu_t } \right)^2 } \mathord{\left/ {\vphantom {{ - \left( {y_t - \mu_t } \right)^2 } {2\sigma^2_t }}} \right. \kern-\nulldelimiterspace} {2\sigma^2_t }}}
\end{equation*}
\\
assuming $f_t = (\mu_t, \sigma^2_t)$. The updating mechanism for time varying parameters $\mu_t$ and $\sigma^2_t$ could be specified as follows:

\begin{equation*}
    f_t = \omega + \mathbf{A} s_{t-1} + \mathbf{B} f_{t-1}
\end{equation*}
\\
where $\theta$ is the vector containing all the scalar parameters in $\omega, \mathbf{A}, \mathbf{B}$ and $s_t$ is scaled by (\ref{eq:6}) setting $\gamma=1$. In particular, the conditional score vectors are in this case given by:

\begin{equation*}
   \nabla_t^{(\mu)} = \frac{(y_t-\mu_t)}{\sigma^2_t} 
\end{equation*}
\begin{equation*}
   \nabla_t^{(\sigma)} = \frac{(y_t-\mu_t)^2}{2\sigma^4_t} - \frac{T}{2\sigma^2_t}
\end{equation*}
\\
Summarizing, the model's variables and parameters are:

\begin{equation*}
	f_t =
	\begin{pmatrix}
	\mu_t \\
	\sigma^2_t
	\end{pmatrix}, \quad \omega =
	\begin{pmatrix}
	\omega_{\mu} \\
	\omega_{\sigma}
	\end{pmatrix} \text{,} \quad \mathbf{A} =
	\begin{pmatrix}
	a_{\mu}  & 0  \\
	0 & a_{\sigma}
	\end{pmatrix} \quad \text{and} \quad \mathbf{B} =
	\begin{pmatrix}
	b_{\mu}  & 0\\
	0 & b_{\sigma}
	\end{pmatrix}
\end{equation*}
\\
In order to estimate conditional moments we consider the following equations for each $k$-th time series:

\begin{equation}
\label{eq:12}
   \mu_t = \omega_{\mu} +  a_{\mu} s_{t-1} + b_{\mu} \mu_{t-1}
\end{equation}
\begin{equation}
\label{eq:13}
    \sigma^2_t = \omega_{\sigma} + a_{\sigma} s_{t-1} + b_{\sigma} \sigma^2_{t-1}
\end{equation}
\\
Once parameters are estimated thorough the usual maximum likelihood approach, conditional moments estimates $\hat{\mu}_t$ and $\hat{\sigma}_t$ are obtained as:

\begin{equation*}
   \hat{\mu}_t = \hat{\omega}_{\mu} + \hat{a}_{\mu} s_{t-1} + \hat{b}_{\mu} \mu_{t-1}
\end{equation*}
\begin{equation*}
    \hat{\sigma}^2_t = \hat{\omega}_{\sigma} + \hat{a}_{\sigma} s_{t-1} + \hat{b}_{\sigma} \sigma^2_{t-1}
\end{equation*}

\subsubsection{t-Dynamic Conditional Score}
\label{tdcs}
As we have seen, stock returns are not normally distributed. In order to account for non-normality we now assume that $y_t$ follows a t-student distribution with location $\mu_t$, scale $\phi_t$ and degrees of freedom $v_t>2$ with its density given by:

\begin{equation*}
     p(y_t| f_t, \mathcal{F}_t; \theta) = \frac{\Gamma\left( \frac{v_t + 1}{2}\right)}{\Gamma\left( \frac{v_t}{2}\right)\phi_t \sqrt{\pi v_t}} \left( 1 + \frac{(y_t - \mu_t)^2}{v_t \phi_t}\right)^{\frac{v_t+1}{2}}
\end{equation*}
\\
Now we have that $f_t = (\mu_t, \phi_t, v_t)$. Assuming a DCS(1,1), the uptading mechanism for $f_t$ is again given by:

\begin{equation*}
    f_t = \omega + \mathbf{A} s_{t-1} + \mathbf{B} f_{t-1}
\end{equation*}
\\
In the case of t-student density, the conditional score vectors are equal to:

\begin{equation*}
    \nabla_t^{(\mu)} = \frac{(v_t + 1)(y_t -\mu_t)}{v_t\phi_t + (y_t-\mu_t)^2}
\end{equation*}
\begin{equation*}
    \nabla_t^{(\phi)} = \frac{1}{2\phi_t} \left[ \frac{(v_t + 1)(y_t -\mu_t)^2}{v_t\phi_t + (y_t-\mu_t)^2} -1 \right]
\end{equation*}
\begin{equation*}
    \nabla_t^{(v)}=\frac{1}{2} \Bigg\{\psi\left( \frac{v_t + 1}{2}\right) - \psi\left( \frac{v_t}{2}\right) - \frac{1}{ v_t} - \log \left( 1 +\frac{(y_t - \mu_t)^2}{v_t \phi_t} \right) + \frac{(v_t+1)(y_t - \mu_t)^2}{v_t \left[ v_t \phi_t + (y_t-\mu_t)^2\right]}\Bigg\}
\end{equation*}
\\
where $\psi(\cdot)$ is the Digamma function. By scaling the conditional score with $\mathcal{I}^{-1}$ in (\ref{eq:6}), the updating mechanism for conditional moments could be written as:

\begin{equation}
\label{eq:14}
   \mu_t = \omega_{\mu} +  a_{\mu} s_{t-1} + b_{\mu} \mu_{t-1}
\end{equation}
\begin{equation}
\label{eq:15}
    \phi_t = \omega_{\phi} + a_{\phi} s_{t-1} + b_{\mu} \phi_{t-1} 
\end{equation}
\begin{equation}
\label{eq:16}
    v_t = \omega_{v} + a_{v} s_{t-1} + b_{v} v_{t-1}
\end{equation}
\\
In compact form:

\begin{equation*}
	f_t =
	\begin{pmatrix}
	\mu_t \\
	\phi_t\\
	v_t
	\end{pmatrix}, \quad \omega =
	\begin{pmatrix}
	\omega_{\mu} \\
	\omega_{\phi}\\
	\omega_{v}
	\end{pmatrix} \text{,} \quad \mathbf{A} =
	\begin{pmatrix}
	a_{\mu}  & 0  & 0\\
	0 & a_{\phi} & 0\\
	0 & 0 & a_{v}\\
	\end{pmatrix} \quad \text{and} \quad \mathbf{B} =
	\begin{pmatrix}
	b_{\mu}  & 0 & 0\\
	0 & b_{\phi} & 0\\
	0 & 0 & b_{v}
	\end{pmatrix}
\end{equation*}
\\
This model has also been defined as Beta-t-EGARCH model by \cite{harvey2013, harvey2014}. To estimate conditional moments, we estimate parameters with MLE and replace them within the conditional moments' equations.

\subsubsection{Skew-t Dynamic Conditional Score}
\label{skewtdcs}
In the last experiment we account for dynamic conditional skewness in asset returns. The density associated to the skew-t distribution is \cite{azzalini2003}:

\begin{equation*}
    p(y_t| f_t, \mathcal{F}_t; \theta) = \frac{2}{\phi_t} \frac{\Gamma\left( \frac{v_t + 1}{2}\right)}{\Gamma\left( \frac{v_t}{2}\right)\sqrt{\pi v_t}} \left( 1 + \frac{(y_t - \mu_t)^2}{v_t \phi_t^2}\right)^{-\frac{v_t+1}{2}} T(y_t \lambda_t \tau_t; v+1) 
\end{equation*}
\\
where the vector of the four time varying parameters is $f_t = (\mu_t, \phi_t, \lambda_t, v_t)$, the term $T$ is given by:

\begin{equation*}
T(z; v) = \int_{-\infty}^{z} t(u;v) du=\int_{-\infty}^{z}
\frac{\Gamma\left(v + 1\right)}{2}{\Gamma\left( \frac{v}{2}\right)\sqrt{\pi v}} \left( 1 + \frac{u^2}{v }\right)^{-\frac{v+1}{2}}du
\end{equation*}
\\
and:

\begin{equation*}
    \tau_t = \Bigg\{\frac{v_t + 1}{\left(\frac{y_t-\mu_t}{\phi_t}\right)^2+v_t}\Bigg\}^{1/2}
\end{equation*}
\\
In such a case, the conditional scores are equal to (see \cite{diciccio2011}):

\begin{equation*}
    \nabla_t^{(\mu)} = \frac{(y_t - \mu_t) \tau^2_t}{\phi_t} - \frac{\lambda_t \tau_t v_t}{\phi_t \left[v_t+(y_t-\mu_t)^2\right]}
\end{equation*}
\begin{equation*}
    \nabla_t^{(\phi)} = -\frac{1}{\phi_t}+\frac{(y_t-\mu_t)^2 \tau^2_t}{\phi_t}-\frac{[\lambda_t \tau_t (y_t-\mu_t)]v_t}{\phi_t[v_t+(y_t-\mu_t)^2]}\phi_t
\end{equation*}
\begin{equation*}
    \nabla_t^{(\lambda)} = (y_t-\mu_t) \tau_t \phi_t
\end{equation*}
\begin{align*}
    \nabla_t^{(v)} &= \frac{1}{2} \Bigg\{\psi\left( \frac{v_t + 1}{2}\right) - \psi\left( \frac{v_t}{2}\right) - \frac{1}{ v_t} - \log \left( 1 +\frac{(y_t - \mu_t)^2}{v_t \phi_t} \right) + \\
    & + \frac{(y_t-\mu_t)^2 \tau^2_t}{v_t \phi_t} + \frac{\lambda_t (y_t-\mu_t)[(y_t-\mu_t)^2-1]}{(v_t+(y_t-\mu_t)^2\tau_t}\phi_t +\frac{\gamma}{T(\lambda_t \tau_t (y_t-\mu_t); v_t +1)} \Bigg\}
\end{align*}
\\
where
\begin{equation*} 
    \gamma = \int_{-\infty}^{\lambda_t \tau_t (y_t-\mu_t)} \Bigg\{ \frac{(v+2)z^2}{(v+1)(v+1+z^2)}-\log \left( 1+ \frac{z^2}{v+1}\right)\Bigg\}t(z;v+1)dz.
\end{equation*}
\\
Written in compact form, we estimate the following skew-t-DCS(1,1) model:

\begin{equation*}
    f_t = \omega + \mathbf{A} s_{t-1} + \mathbf{B} f_{t-1}
\end{equation*}
\\
where now we have:

\begin{equation*}
	f_t =
	\begin{pmatrix}
	\mu_t \\
	\phi_t\\
	\lambda^2_t\\
	v_t
	\end{pmatrix}, \quad \omega =
	\begin{pmatrix}
	\omega_{\mu} \\
	\omega_{\phi}\\
	\omega_{\lambda}\\
	\omega_{v}
	\end{pmatrix} \text{,} \quad \mathbf{A} =
	\begin{pmatrix}
	a_{\mu}  & 0  & 0 & 0\\
	0 & a_{\phi} & 0 & 0\\
	0 & 0 & a_{\lambda} & 0\\
	0 & 0 & 0 & a_{v}\\
	\end{pmatrix} \quad \text{and} \quad \mathbf{B} =
	\begin{pmatrix}
	b_{\mu} & 0 & 0  & 0\\
	0 & b_{\phi} & 0 & 0\\
	0 & 0 & b_{\lambda}&0\\
	0 & 0 & 0 & b_{v}
	\end{pmatrix}
\end{equation*}
\\
The conditional moments equation could be written as:

\begin{equation*}
   \mu_t = \omega_{\mu} +  a_{\mu} s_{t-1} + b_{\mu} \mu_{t-1}
\end{equation*}
\begin{equation*}
    \phi_t = \omega_{\phi} + a_{\phi} s_{t-1} + b_{\mu} \phi_{t-1} 
\end{equation*}
\begin{equation}
\label{eq:17}
    \lambda_t = \omega_{\lambda} + a_{\lambda} s_{t-1} + b_{\lambda} \lambda_{t-1}
\end{equation}
\begin{equation*}
    v_t = \omega_{v} + a_{v} s_{t-1} + b_{v} v_{t-1}
\end{equation*}
\\
Where the third one allows for time varying skewness. Replacing MLE estimates $\hat{\omega}, \hat{\mathbf{A}}, \hat{\mathbf{B}}$ we get estimated conditional moments.

\section{Empirical experiments: results}
\label{sec5}

In what follows, we provide a detailed explanation of clustering results.

\subsection{Fuzzy clustering under normality assumption}

The first step is to estimate the model presented in the Section \ref{guassiandcs}. 
%
%

Conditional moments time series are shown, for the sake of brevity, within the Appendix section. Before the identification of the clusters through the application of the A-FCM clustering algorithm of \cite{durso2009}, we have found the optimal number of clusters. Following a fuzzy approach, in the Tab. \ref{tab:t2} are reported the values of the fuzzy silhouette criterion proposed by \cite{campello2006}. The optimal number of clusters is the one with the highest value.

\begin{table}[H]
    \centering
    \begin{tabular}{ccc}
    \hline
         No. Clusters & $\hat{\mu}_t$ & $\hat{\sigma}^2_t$ \\
         \hline
        2 & 0.9814 & 0.7846 \\
        3 & 0.9480 & 0.6418 \\
        4 & 0.8852 & 0.6303 \\
        5 & 0.4559 & 0.6333 \\
        \hline
    \end{tabular}
    \caption{Fuzzy silhouette of \cite{campello2006}}
\label{tab:t2}
\end{table}

Therefore, we have $\mathcal{C}=2$ number of clusters for all $R=2$ conditional moments. In the first, we have two groups with the same conditional mean while in the second we have two groups with the same conditional variance. Particularly the DCS-AFCM procedure results in the following assignment:

\begin{align*}
     \textit{Mean} \quad  & \text{\textbf{Group 1}: (AA, AXP, BA, C, CAT, CVX, DD, DIS, }\\
   & \text{HD, HPQ IBM, INTC, JNJ, JPM, KO, MMM, MRK, } \\
    & \text{MSFT, PFE, PG, T, UTX, VZ, WMT, XOM);}\\
    & \text{\textbf{Group 2}: (BAC, GE, GM, AIG).}\\
\\
   \textit{Variance} \quad  & \text{\textbf{Group 1}: (AXP, CAT, CVX, GE, KO, MMM, MRK);}\\
    & \text{\textbf{Group 2}: (AA, BA, BAC, C, DD, DIS, GM, HD,}\\
    & \text{HPQ, IBM, INYC, JNJ, JPM, AIG, MCD, MSFT,}\\
    & \text{PFE, PG, T, UTX, VZ, WMT, XOM).}
\end{align*}
\\
Further, in the Tab. \ref{tab:t3} we reported the membership degrees of the assignment or, alternatively, the level of probability that each time series has to be in each $\mathcal{C}=2$ clusters. In the last columns, instead, are reported the final assignment.

\begin{table}[H]
\centering
\begin{tabular}{rrrr|rrr}
  \hline
  \multicolumn{1}{c}{} &
\multicolumn{2}{c}{Mean} &
\multicolumn{1}{c}{} & 
\multicolumn{2}{c}{Variance} &
\multicolumn{1}{c}{}\\
 Stock & Cluster 1 & Cluster 2 & Assigned & Cluster 1 & Cluster 2 & Assigned \\ 
  \hline
AA & 0.9980 & 0.0020 & 1 & 0.4546 & 0.5454 & 2 \\
  AIG & 0.0010 & 0.9990 & 2 & 0.4622 & 0.5378 & 2\\ 
  AXP & 0.9991 & 0.0009 & 1 & 0.6035 & 0.3965 & 1\\ 
  BA & 0.9986 & 0.0014  & 1 & 0.0185 & 0.9815 & 2 \\ 
  BAC & 0.0008 & 0.9992 & 2 & 0.0961 & 0.9039 & 2\\ 
  C & 0.9965 & 0.0035 & 1 & 0.0227 & 0.9773 & 2\\ 
  CAT & 0.9943 & 0.0057 & 1 & 0.9622 & 0.0378 & 1\\ 
  CVX & 0.9991 & 0.0009 & 1  & 0.6756 & 0.3244 & 1\\ 
  DD & 0.9993 & 0.0007 & 1 & 0.0543 & 0.9457 & 2\\ 
  DIS & 0.9995 & 0.0005 & 1 & 0.1923 & 0.8077 & 2\\ 
  GE & 0.0084 & 0.9916 & 2 & 0.9846 & 0.0154 & 1 \\ 
  GM & 0.0027 & 0.9973 & 2 & 0.0853 & 0.9147 & 2 \\ 
  HD & 0.9989 & 0.0011 & 1 & 0.3302 & 0.6698 & 2\\ 
  HPQ & 0.9914 & 0.0086 & 1 & 0.0266 & 0.9734 & 2\\ 
  IBM & 0.6448 & 0.3552 & 1 & 0.2403 & 0.7597 & 2\\ 
  INTC & 0.9983 & 0.0017 & 1 & 0.0276 & 0.9724 & 2\\ 
  JNJ & 0.9984 & 0.0016 & 1 & 0.2611 & 0.7389 & 2\\ 
  JPM & 0.9987 & 0.0013 & 1 & 0.0200 & 0.9800 & 2\\ 
  KO & 0.9993 & 0.0007 & 1 & 0.7768 & 0.2232 & 1\\ 
  MCD & 0.9992 & 0.0008 & 1 & 0.0607 & 0.9393 & 2\\ 
  MMM & 0.9990 & 0.0010 & 1 & 0.9520 & 0.0480 & 1\\ 
  MRK & 0.9991 & 0.0009 & 1  & 0.9257 & 0.0743 & 1\\ 
  MSFT & 0.9984 & 0.0016 & 1  & 0.0039 & 0.9961 & 2 \\ 
  PFE & 0.9989 & 0.0011 & 1 & 0.0189 & 0.9811 & 2\\ 
  PG & 0.9977 & 0.0023 & 1 & 0.0176 & 0.9824 & 2\\ 
  T & 0.9988 & 0.0012 & 1 & 0.0483 & 0.9517 & 2\\ 
  UTX & 0.9992 & 0.0008 & 1 & 0.0458 & 0.9542 & 2\\ 
  VZ & 0.9988 & 0.0012 & 1 & 0.0691 & 0.9309 & 2\\ 
  WMT & 0.9851 & 0.0149 & 1 & 0.1435 & 0.8565 & 2\\ 
  XOM & 0.9971 & 0.0029 & 1 & 0.3379 & 0.6621 & 2\\ 
   \hline
\end{tabular}
\caption{Clustering based on conditional mean (left) and variance (right) estimates. Membership degrees for each cluster are reported, while the group's assignment (\emph{assigned}) after the DCS-based AFCM procedure is in the last column.}
\label{tab:t3}
\end{table}

The 86.7\% of the assets are clustered in the same group about conditional mean estimated trough a Gaussian Dynamic Conditional Score model while, as the other side, 76.7\% of them are within the second group according to the conditional variances' pattern. Moreover, cluster assignment is much more uncertain for some assets when the conditional variance is considered, rather than the conditional mean. For example, some assets like \textit{AA} or \textit{AIG} have almost the same membership degree for both clusters 1 and 2 when we take the conditional variance. This never happens for conditional mean clustering. Therefore, here we have a clear example of the fuzzy clustering importance; we could be able to assign asset like \textit{AA} and \textit{AIG} to both clusters with almost the same probability.
\\
Moreover, starting from these $R=2$ levels of clusters, we are also able to group time series with exactly the same conditional distribution over time. This could be done by checking for assets that are within the same group for both conditional mean and variance. Particularly, in this case, the assets \textit{BAC, GM} and \textit{AIG} are always in the second group, so we say that they share the same conditional distribution assuming only two time-varying moments. On the other side, the assets \textit{AXP, CAT, CVX, KO, MMM} and \textit{MRK} are always in the first group according to both moments. Hence, we could generate an additional cluster with those assets that share the same conditional distribution. Nevertheless, those clusters could be formed according to certain probabilities. Suppose we assign, about conditional variance, with the 46\% of probability both \textit{AA} and \textit{AIG} to the cluster 1 instead of 2. In this case \textit{AIG} does not have anymore the same conditional distribution of \textit{BAC} and \textit{GM}, while \textit{AA} will have the same conditional distribution of  \textit{AXP, CAT, CVX, KO, MMM} and \textit{MRK} with the 46\% of probability.

\subsection{Fuzzy clustering under t-student density}

Considering stock retuns' non normality, we estimated the $t-DCS(1,1)$ model presented in the Section \ref{tdcs}. 
%
%
The optimal number of clusters has been decided according to the fuzzy silhouette criterion whose results are showed in the Tab. \ref{tab:t7}.

\begin{table}[H]
    \centering
    \begin{tabular}{cccc}
    \hline
         No. Clusters & $\hat{\mu}_t$ & $\hat{\phi}_t$ & $\hat{v}_t$ \\
         \hline
        2 & 0.9974 & 0.8281 & 0.8301 \\
        3 & 0.9924 & 0.7703 & 0.7010\\
        4 & 0.9559 & 0.7621 & 0.6988\\
        5 & 0.9600 & 0.6887 & 0.4260\\
        \hline
    \end{tabular}
    \caption{Fuzzy silhouette of \cite{campello2006}}
\label{tab:t7}
\end{table}

The clusters obtained according to the t-DCS-AFCM procedure are the following:

\begin{align*}
    \textit{Location} \quad  & \text{\textbf{Group 1}: (C, HPQ, INTC, JPM);}\\
    & \text{\textbf{Group 2}: (AA, AXP, BA, BAC, CAT, CVX, DD, DIS,}\\
    & \text{GE, GM, HD, IBM, JNJ, AIG, KO, MCD, MMM, MRK,}\\
    & \text{MSFT, PFE, PG, T, UTX, VZ, WMT, XOM).}\\
\\
    \textit{Scale} \quad  & \text{\textbf{Group 1}: (AA, BA, CAT, CVX, DD, DIS, GE, GM,}\\
    & \text{AIG, MRK, PFE, T, VZ);}\\\
    & \text{\textbf{Group 2}: (AXP, BAC, C, HD, HPQ, IBM, INTC,}\\
    & \text{JNJ, JPM, KO, MCD, MMM, MSFT, PG, VTZ, WMT).}\\
\\
    \textit{Shape} \quad  & \text{\textbf{Group 1}: (BAM CAT, DD, HD, HPQ, IBM, INTC, JNJ,}\\
  & \text{AIG, KO, MCD, MMM, MRK, MSFT, OFE, PG, T, }\\
   & \text{UTX, VZ, WMT, XOM);}\\\
    & \text{\textbf{Group 2}: (AA, AXP, BAC, C, CVX, DIS, GE, GM, JPM).}
    \end{align*}

Moreover, in the Tab. \ref{tab:t8} we reported the uncertainty about clustering assignment in such scenario. Differently from Gaussian scenario, here we have $R=3$ levels of clusters. In the first, we have two groups with the same conditional location, in the second we have two groups with the same conditional scale, and in the last one we have two groups with the same conditional shape.
\\
\\
In the case, we have, as the previous one, almost no uncertainty in clustering time series according to their first conditional moment (see Tab. \ref{tab:t8}). Instead, a bit of uncertainty rise in clustering two assets (\textit{BAC} and \textit{T}) about their second conditional moment (Tab. \ref{tab:t8}). Instead, we have also low uncertainty in clustering time series according to conditional shape (Tab. \ref{tab:t8}). About clusters composition, the first level has the 86.7\% of the assets within the same group for conditional mean, even if now most of them are located in the group 2 instead of 1. The cluster level $r=2$ is more balanced since the 43.3\% of the assets are in the first group and the rest in the second one. At this point, we have that most of the assets are located in the same group. Indeed, we have that 11 of the 16 assets within the same cluster for conditional scale are also in the same group for the conditional mean. Therefore, we could conclude that 11 assets (\textit{AXP, BAC, HD, IBM, JNJ, KO, MCD, MMM, MSFT, PG, VTX, WMT}) are similar in terms of conditional location and scale. Nevertheless, not all of them share the same conditional distribution. Indeed, only \textit{AXP} and \textit{BAC} are always in the same group, and we say that they share a similar conditional distribution. However, we have to highlight a probabilistic argument that is very important is such a scenario. Indeed, we should account for the uncertainty about the conditional second-moment group assignment for the stock \textit{BAC}. With a probability of 45\% its conditional scale could be assigned to group 1 instead of 2. In this case, with such level of probability, no stock could be considered with a similar conditional distribution. 

\begin{table}[H]
\centering
\scalebox{0.75}{\begin{tabular}{rrrrrrrrrr}
 \hline
\multicolumn{1}{c}{} &
\multicolumn{2}{c}{Location} &
\multicolumn{1}{c}{} &
\multicolumn{2}{c}{Scale} &
\multicolumn{1}{c}{} &
\multicolumn{2}{c}{Shape} &
\multicolumn{1}{c}{}\\
 & Cluster 1 & Cluster 2 & Assigned & Cluster 1 & Cluster 2 & Assigned & Cluster 1 & Cluster 2 & Assigned\\ 
  \hline
AA & 0.0009 & 0.9991 & 2 & 0.9260 & 0.0740 & 1 & 0.1429 & 0.8571 & 2\\
 AIG & 0.0013 & 0.9987& 2 & 0.9854 & 0.0146 & 1 & 0.9257 & 0.0743 & 1\\ 
  AXP & 0.0004 & 0.9996 & 2 & 0.1549 & 0.8451 & 2 & 0.0151 & 0.9849 & 2\\ 
  BA & 0.0010 & 0.9990 & 2 & 0.9562 & 0.0438 & 1 & 0.9868 & 0.0132 & 1\\ 
  BAC & 0.0011 & 0.9989 & 2 & 0.4525 & 0.5475 & 2 & 0.1399 & 0.8601 & 2\\ 
  C & 0.9981 & 0.0019 & 1 & 0.0062 & 0.9938 & 2 & 0.0492 & 0.9508 & 2\\ 
  CAT & 0.0006 & 0.9994 & 2 & 0.9259 & 0.0741 & 1 & 0.8152 & 0.1848 & 1\\ 
  CVX & 0.0003 & 0.9997 & 2 & 0.9539 & 0.0461 & 1 & 0.9475 & 0.0525 & 1\\ 
  DD & 0.0052 & 0.9948 & 2 & 0.9601 & 0.0399 & 1 & 0.9580 & 0.0420 & 1\\ 
  DIS & 0.0004 & 0.9996 & 2 & 0.9770 & 0.0230 & 1 & 0.2357 & 0.7643 & 2 \\ 
  GE & 0.0002 & 0.9998 & 2 & 0.8373 & 0.1627 & 1 & 0.0206 & 0.9794 & 2\\ 
  GM & 0.0030 & 0.9970 & 2 & 0.9074 & 0.0926 & 1 & 0.0458 & 0.9542 & 2\\ 
  HD & 0.0005 & 0.9995 & 2 & 0.0028 & 0.9972 & 2 & 0.9639 & 0.0361 & 1\\ 
  HPQ & 0.9975 & 0.0025 & 1 & 0.1095 & 0.8905 & 2 & 0.8318 & 0.1682 & 1\\ 
  IBM & 0.0009 & 0.9991 & 2 & 0.0230 & 0.9770 & 2 & 0.9752 & 0.0248 & 1\\ 
  INTC & 0.9966 & 0.0034 & 1 & 0.0166 & 0.9834 & 2 & 0.9249 & 0.0751 & 1\\ 
  JNJ & 0.0008 & 0.9992 & 2 & 0.0126 & 0.9874 & 2 & 0.9565 & 0.0435 & 1\\ 
  JPM & 0.9981 & 0.0019 & 1 & 0.0476 & 0.9524 & 2 & 0.0112 & 0.9888 & 2\\ 
  KO & 0.0006 & 0.9994 & 2 & 0.0042 & 0.9958 & 2 & 0.9746 & 0.0254 & 1\\ 
  MCD & 0.0005 & 0.9995 & 2 & 0.1480 & 0.8520 & 2  & 0.8663 & 0.1337 & 1\\ 
  MMM & 0.0006 & 0.9994 & 2 & 0.0465 & 0.9535 & 2 & 0.9578 & 0.0422 & 1\\ 
  MRK & 0.0006 & 0.9994 & 2 & 0.9392 & 0.0608 & 1 & 0.9636 & 0.0364 & 1\\ 
  MSFT & 0.0007 & 0.9993 & 2 & 0.0949 & 0.9051 & 2 & 0.9875 & 0.0125 & 1\\ 
  PFE & 0.0006 & 0.9994 & 2 & 0.5089 & 0.4911 & 1 & 0.8743 & 0.1257 & 1\\ 
  PG & 0.0005 & 0.9995 & 2 & 0.0065 & 0.9935 & 2 & 0.9794 & 0.0206 & 1\\ 
  T & 0.0015 & 0.9985 & 2 & 0.5479 & 0.4521 & 1 & 0.8519 & 0.1481 & 1\\ 
  UTX & 0.0004 & 0.9996 & 2 & 0.2354 & 0.7646 & 2 & 0.8805 & 0.1195 & 1\\ 
  VZ & 0.0006 & 0.9994 & 2 & 0.9498 & 0.0502 & 1 & 0.9101 & 0.0899 & 1\\ 
  WMT & 0.0011 & 0.9989 & 2 & 0.1892 & 0.8108 & 2 & 0.8059 & 0.1941 & 1\\ 
  XOM & 0.0010 & 0.9990 & 2 & 0.9379 & 0.0621 & 1 & 0.9811 & 0.0189 & 1 \\
   \hline
\end{tabular}}
\caption{Clustering based on conditional location (left), scale (center) and shape (right) estimates. Membership degrees for each cluster are reported, while the group's assignment (\emph{assigned}) after the DCS-based AFCM procedure is in the last columns.}
\label{tab:t8}
\end{table}

\subsection{Fuzzy clustering under skew-t density}

The last experiment allows for time-varying skewness with the additional equation (\ref{eq:17}) that has to be estimated. 
%
%

Then we cluster time series according to their moments' estimate via the skew-t-DCS model. Optimal number of clusters is shown in Tab. \ref{tab:t12}.

\begin{table}[H]
    \centering
    \begin{tabular}{ccccc}
    \hline
         No. Clusters & $\hat{\mu}_t$ & $\hat{\phi}_t$  & $\hat{\lambda}_t$ & $\hat{v}_t$ \\
         \hline
        2 & 0.8979 & 0.7113 & 0.8629 & 0.8321 \\
        3 & 0.9408 & 0.6410 & 0.8563 & 0.8408 \\
        4 & 0.8409 & 0.5708 & 0.8404 & 0.8573 \\
        5 & 0.8690 & 0.5314 & 0.8353 & 0.8798 \\
        \hline
    \end{tabular}
    \caption{Fuzzy silhouette of \cite{campello2006}}
    \label{tab:t12}
\end{table}

With this last experiment we do not get the same optimal number of clusters for all considered higher conditional moments. Indeed, while we consider $\mathcal{C}=3$ clusters for conditional location, we have $\mathcal{C}=2$ for both conditional scale and skewness and $\mathcal{C}=5$ for conditional shape. In this case, it is evidently more difficult than in the previous models to have time series within the same clusters for all conditional moments. 
The results of the proposed skew-t-DCS model-based A-FCM algorithm are summarized by the following scheme:

\begin{align*}
    \textit{Location} \quad  & \text{\textbf{Group 1}: (BAC, JPM);}\\
    & \text{\textbf{Group 2}: (AA, AXP, BA, C, CAT, CVX, DD, DIS,}\\
    & \text{GE, GM, INTC, KO, MCD, MMM, MRK, PFE,}\\
    & \text{PG, T, UTX, WMT, XOM);}\\
     & \text{\textbf{Group 3}: (HD, HPQ, IBM, JNJ, AIG, MSFT, VZ).}\\
\\
   \textit{Scale} \quad  & \text{\textbf{Group 1}: (AA, AXP, BA, BAC, CAT, DD, GE, GM,}\\
   & \text{HD, HPQ, INTC, IBM, JPM, MCD, MMM, PFE,}\\
    & \text{T, UTX, VZ, WMT);}\\
    & \text{\textbf{Group 2}: (C, CVX, DIS, JNJ, AIG, KO, MRK,}\\
    & \text{MSFT, PG, XOM).}\\
\\
    \textit{Skewness} \quad  & \text{\textbf{Group 1}: (BAC, CVX, DD, GE, GM, INTC, JPM, AIG,}\\
    & \text{MCD, MMM, MRK, MSFT, PG, T, UTX, VZ, WMT, XOM);}\\
    & \text{\textbf{Group 2}: (AA, AXP, BA, C, CAT, CVX, DIS, HD}\\
    & \text{HPQ, IBM, JNJ, KO, PFE).}\\
\\
    \textit{Shape} \quad  & \text{\textbf{Group 1}: (AA, BAC, DIS, GE, KO, MCD, PG, WMT);}\\
   & \text{\textbf{Group 2}: (AXP, CAT, CVX, DD, GM, HD, HPQ,}\\
    & \text{INTC, MMM, MRK, MSFT);}\\
    & \text{\textbf{Group 3}: (BA, C, PFE, T, VZ);}\\
    & \text{\textbf{Group 4}: (IBM, JPM, UTX);}\\
    & \text{\textbf{Group 5}: (JNJ, AIG, XOM);}
\end{align*}
\\
About clusters composition, only 6.7\% of the assets are in the first group for conditional mean, while the second group contains the majority of the assets. About conditional scale, the first group is the one with the highest number of assets (66.7\%) and the same apply for conditional skewness, where there are the 60\% of the assets. About conditional shape, the group 3, 4 and 5 are the ones with lower assets (13\%, 10\% and 10\%, respectively), while the group 2 is the most numerous (36.7\%).
\\
Once again, clustering made according to conditional mean is the one with the lowest level of uncertainty since the membership degree of any asset to the alternative groups is tiny (Tab. \ref{tab:t13}). Differently, as we have also seen for the other two experiments, the clusters based on the conditional scale are the ones with the highest level of uncertainty (Tab. \ref{tab:t14}). For example, some assets -- like \textit{PFE} and \textit{BAC} -- have 35\% probability to be clustered in a different group; for \textit{PG} such a probability is 40\%, while others like \textit{C} and \textit{BA} also a higher probability of 47\% to be in another cluster. In the end, while also the conditional skewness-based clusters show some relevant cases of uncertainty (e.g. \textit{HPQ, JPM, MRK, T} or \textit{UTX}); differently, shape-based clusters present a low level of uncertainty (see Tab. \ref{tab:t15} and Tab. \ref{tab:t16}).
\\
Overall, while we can build $R$ clustering levels, in this context, accounting for cluster uncertainty evident for some conditional moments like scale and skewness, we are not able to find stocks always within the same clusters. In other words, all stocks have different conditional distributions regarding higher moments. If we look at just some of them, we could easily get in error.

\begin{table}[H]
\centering
\begin{tabular}{rrrr|r}
  \hline
 & Cluster 1 & Cluster 2 & Cluster 3 & Assignment \\ 
  \hline
AA & 0.0290 & 0.8263 & 0.1447 & 2\\ 
  AIG & 0.0257 & 0.0615 & 0.9128 & 3\\ 
  AXP & 0.0005 & 0.9978 & 0.0018 & 2\\ 
  BA & 0.0008 & 0.9961 & 0.0031 & 2\\ 
  BAC & 0.9632 & 0.0105 & 0.0263 & 1\\ 
  C & 0.0010 & 0.9954 & 0.0036 & 2\\ 
  CAT & 0.0016 & 0.9927 & 0.0057 & 2\\ 
  CVX & 0.0020 & 0.9906 & 0.0073 & 2\\ 
  DD & 0.0005 & 0.9976 & 0.0019 & 2\\ 
  DIS & 0.0091 & 0.9522 & 0.0387 & 2 \\ 
  GE & 0.0082 & 0.9568 & 0.0350 & 2\\ 
  GM & 0.0023 & 0.9896 & 0.0082 & 2\\ 
  HD & 0.0017 & 0.0022 & 0.9961 & 3\\ 
  HPQ & 0.0060 & 0.0063 & 0.9877 & 3\\ 
  IBM & 0.0157 & 0.0135 & 0.9709 & 3\\ 
  INTC & 0.0012 & 0.9944 & 0.0044 & 2\\ 
  JNJ & 0.1477 & 0.0572 & 0.7952 & 3\\ 
  JPM & 0.9393 & 0.0131 & 0.0476 & 1\\ 
  KO & 0.0036 & 0.9817 & 0.0146 & 2\\ 
  MCD & 0.0022 & 0.9898 & 0.0079 & 2\\ 
  MMM & 0.0053 & 0.9736 & 0.0211 & 2\\ 
  MRK & 0.0023 & 0.9886 & 0.0091 & 2\\ 
  MSFT & 0.0034 & 0.0043 & 0.9923 & 3\\ 
  PFE & 0.0021 & 0.9901 & 0.0077 & 2\\ 
  PG & 0.0006 & 0.9970 & 0.0024 & 2\\ 
  T & 0.0020 & 0.9907 & 0.0073 & 2\\ 
  UTX & 0.0020 & 0.9910 & 0.0070 & 2\\ 
  VZ & 0.0444 & 0.1828 & 0.7728 & 3\\ 
  WMT & 0.0014 & 0.9934 & 0.0051 & 2 \\ 
  XOM & 0.0031 & 0.9861 & 0.0108 & 2\\ 
   \hline
\end{tabular}
\caption{Clustering based on conditional location estimates. Membership degrees for each cluster are reported, while the group's assignment after the DCS-based AFCM procedure is in the last column.}
\label{tab:t13}
\end{table}

\begin{table}[H]
\centering
\begin{tabular}{rrr|r}
  \hline
 & Cluster 1 & Cluster 2 & Assignment \\ 
  \hline
AA & 0.7663 & 0.2337 & 1\\ 
  AIG & 0.0720 & 0.9280 & 2\\ 
  AXP & 0.7140 & 0.2860 & 1\\ 
  BA & 0.5230 & 0.4770 & 1\\ 
  BAC & 0.6448 & 0.3552 & 1\\ 
  C & 0.4727 & 0.5273 & 2\\ 
  CAT & 0.9703 & 0.0297 & 1\\ 
  CVX & 0.0996 & 0.9004 & 2\\ 
  DD & 0.9130 & 0.0870 & 1\\ 
  DIS & 0.3350 & 0.6650 & 2\\ 
  GE & 0.9820 & 0.0180 & 1\\ 
  GM & 0.9716 & 0.0284 & 1\\ 
  HD & 0.9351 & 0.0649 & 1\\ 
  HPQ & 0.8275 & 0.1725 & 1\\ 
  IBM & 0.9752 & 0.0248 & 1\\ 
  INTC & 0.9883 & 0.0117 & 1\\ 
  JNJ & 0.0155 & 0.9845 & 2\\ 
  JPM & 0.8707 & 0.1293 & 1\\ 
  KO & 0.0854 & 0.9146 & 2\\ 
  MCD & 0.7563 & 0.2437 & 1\\ 
  MMM & 0.9037 & 0.0963 & 1 \\ 
  MRK & 0.0534 & 0.9466 & 2\\ 
  MSFT & 0.0684 & 0.9316 & 2\\ 
  PFE & 0.6454 & 0.3546 & 1\\ 
  PG & 0.3903 & 0.6097 & 2\\ 
  T & 0.9899 & 0.0101 & 1\\ 
  UTX & 0.6893 & 0.3107 & 1\\ 
  VZ & 0.8634 & 0.1366 & 1\\ 
  WMT & 0.8690 & 0.1310 & 1 \\ 
  XOM & 0.1527 & 0.8473 & 2\\ 
   \hline
\end{tabular}
\caption{Clustering based on conditional scale estimates. Membership degrees for each cluster are reported, while the group's assignment after the DCS-based AFCM procedure is in the last column.}
\label{tab:t14}
\end{table}

\begin{table}[H]
\centering
\begin{tabular}{rrr|r}
  \hline
 & Cluster 1 & Cluster 2 & Assignment \\ 
  \hline
AA & 0.0389 & 0.9611 & 2\\ 
 AIG & 0.8938 & 0.1062 & 1\\ 
  AXP & 0.0746 & 0.9254 & 2\\ 
  BA & 0.2149 & 0.7851 & 2\\ 
  BAC & 0.9888 & 0.0112 & 1\\ 
  C & 0.0418 & 0.9582 & 2\\ 
  CAT & 0.0219 & 0.9781 & 2\\ 
  CVX & 0.9768 & 0.0232 & 1\\ 
  DD & 0.9885 & 0.0115 & 1\\ 
  DIS & 0.0098 & 0.9902 & 2 \\ 
  GE & 0.9791 & 0.0209 & 1\\ 
  GM & 0.9325 & 0.0675 & 1 \\ 
  HD & 0.0309 & 0.9691 & 2\\ 
  HPQ & 0.4540 & 0.5460 & 2\\ 
  IBM & 0.0344 & 0.9656 & 2\\ 
  INTC & 0.9708 & 0.0292 & 1 \\ 
  JNJ & 0.9932 & 0.0068 & 1\\ 
  JPM & 0.6408 & 0.3592 & 1\\ 
  KO & 0.0227 & 0.9773 & 2\\ 
  MCD & 0.9744 & 0.0256 & 1\\ 
  MMM & 0.9803 & 0.0197 & 1\\ 
  MRK & 0.5795 & 0.4205 & 1\\ 
  MSFT & 0.9941 & 0.0059 & 1 \\ 
  PFE & 0.0368 & 0.9632 & 2\\ 
  PG & 0.9491 & 0.0509 & 1\\ 
  T & 0.5896 & 0.4104 & 1\\ 
  UTX & 0.6041 & 0.3959 & 1\\ 
  VZ & 0.9860 & 0.0140 & 1\\ 
  WMT & 0.9468 & 0.0532 & 1\\ 
  XOM & 0.9988 & 0.0012 & 1\\ 
   \hline
\end{tabular}
\caption{Clustering based on conditional skewness estimates. Membership degrees for each cluster are reported, while the group's assignment after the DCS-based AFCM procedure is in the last column.}
\label{tab:t15}
\end{table}

\begin{table}[H]
\centering
\begin{tabular}{rrrrrr|r}
  \hline
 & Cluster 1 & Cluster 2 & Cluster 3 & Cluster 4 & Cluster 5 & Assignment \\
  \hline
AA & 0.7930 & 0.0611 & 0.0096 & 0.0253 & 0.1110 & 1\\
  AIG & 0.0844 & 0.0156 & 0.0205 & 0.1091 & 0.7705 & 5\\ 
  AXP & 0.0055 & 0.9900 & 0.0007 & 0.0012 & 0.0026 & 2\\ 
  BA & 0.0167 & 0.0079 & 0.8724 & 0.0729 & 0.0300 & 3\\ 
  BAC & 0.8868 & 0.0383 & 0.0050 & 0.0140 & 0.0558 & 1\\ 
  C & 0.0011 & 0.0006 & 0.9925 & 0.0039 & 0.0019 & 3\\ 
  CAT & 0.0273 & 0.9485 & 0.0038 & 0.0068 & 0.0137 & 2\\ 
  CVX & 0.0007 & 0.9988 & 0.0001 & 0.0001 & 0.0003 & 2\\ 
  DD & 0.2191 & 0.6747 & 0.0120 & 0.0262 & 0.0680 & 2 \\ 
  DIS & 0.7352 & 0.0198 & 0.0069 & 0.0266 & 0.2115 & 1\\ 
  GE & 0.8915 & 0.0108 & 0.0032 & 0.0114 & 0.0830 & 1\\ 
  GM & 0.0141 & 0.9740 & 0.0018 & 0.0033 & 0.0068 & 2\\ 
  HD & 0.2535 & 0.6286 & 0.0130 & 0.0288 & 0.0760 & 2\\ 
  HPQ & 0.2094 & 0.6876 & 0.0117 & 0.0255 & 0.0658 & 2\\ 
  IBM & 0.0029 & 0.0009 & 0.0028 & 0.9846 & 0.0089 & 4 \\ 
  INTC & 0.0057 & 0.9903 & 0.0006 & 0.0011 & 0.0024 & 2\\ 
  JNJ & 0.0581 & 0.0058 & 0.0039 & 0.0179 & 0.9143 & 5\\ 
  JPM & 0.0019 & 0.0006 & 0.0020 & 0.9903 & 0.0053 & 4\\ 
  KO & 0.7283 & 0.1328 & 0.0112 & 0.0288 & 0.0989 & 1\\ 
  MCD & 0.8255 & 0.0158 & 0.0050 & 0.0186 & 0.1351 & 1\\ 
  MMM & 0.2048 & 0.6938 & 0.0115 & 0.0251 & 0.0647 & 2\\ 
  MRK & 0.0466 & 0.9095 & 0.0071 & 0.0125 & 0.0243 & 2\\ 
  MSFT & 0.0148 & 0.9726 & 0.0019 & 0.0035 & 0.0072 & 2\\ 
  PFE & 0.0105 & 0.0055 & 0.9368 & 0.0301 & 0.0170 & 3\\ 
  PG & 0.8865 & 0.0117 & 0.0035 & 0.0125 & 0.0858 & 1\\ 
  T & 0.0144 & 0.0076 & 0.9146 & 0.0403 & 0.0231 & 3\\ 
  UTX & 0.0039 & 0.0012 & 0.0052 & 0.9788 & 0.0109 & 4\\ 
  VZ & 0.0114 & 0.0054 & 0.9149 & 0.0479 & 0.0203 & 3\\ 
  WMT & 0.7092 & 0.0190 & 0.0069 & 0.0262 & 0.2386 & 1\\ 
  XOM & 0.1051 & 0.0165 & 0.0166 & 0.1344 & 0.7275 & 5\\ 
   \hline
\end{tabular}
\caption{Clustering based on conditional shape estimates. Membership degrees for each cluster are reported, while the group's assignment after the DCS-based AFCM procedure is in the last column.}
\label{tab:t16}
\end{table}

\section{Experiments with simulated data}

\begin{figure}[H]
    \centering
    \includegraphics[width=\textwidth]{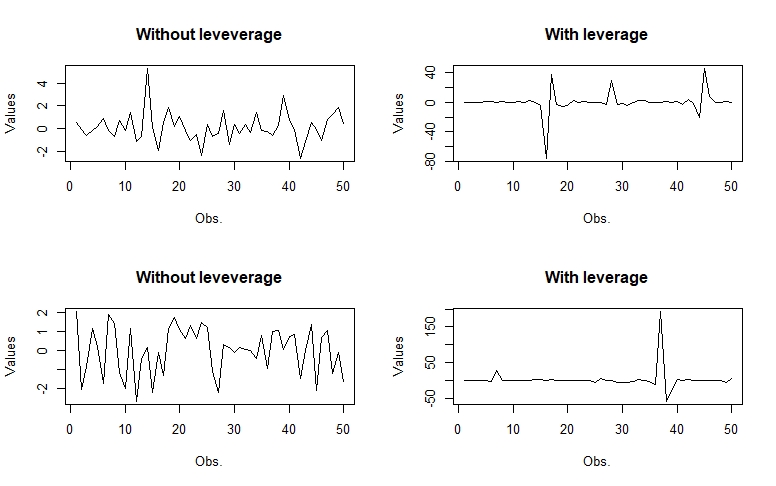}
    \caption{Simulated time series from different D.G.P. (\ref{eq:17}) with and without leverage.}
    \label{fig:sim1}
\end{figure}

As robustness study, in this section we illustrate the results of the proposed clustering procedure with simulated data. At this aim, we generate $N=10$ independent time series of different lengths $T=50, 200, 1000$ belonging to different Data Generating Processes (D.G.P.). The consideration of different lengths is useful to evaluate the performances of the proposed algorithm for either short or long time series. 
\\
Moreover, to evaluate how the selection of the number of lags $L$ affects the results, we consider as in \cite{diaz2010} different choice $L=10, 25, 50$. The performance of the proposed clustering procedure is evaluated, following \cite{diaz2010}, as the correct classification rate over $M=300$ replications. 
\\
We suppose the presence of $\mathcal{C}=2$ clusters with $N=5$ time series in each group, where the $N=10$ time series are generated by the following D.G.P.: 

\begin{align}
    y_{k,t} &= \exp \left(\eta_{k,t} \right)\epsilon_{k,t}=\sigma_{k,t}\epsilon_{k,t} \quad \epsilon_t \sim st(0,\sigma^2_{\epsilon}, \lambda, v) \nonumber\\
   \eta_{k,t} &= \omega + \phi \eta_{k,t-1}  + \alpha u_{k,t-1} + \beta \text{sgn}(-y_{k,t-1})(u_{k,t-1}+1)
\label{eq:17}
\end{align}
\\
that is the Beta-Skew-t-EGARCH of \cite{harvey2014}, hence a non-Gaussian and highly non-linear process, with $\omega$ be the log-scale intercept, $\phi$ is the persistency parameter, $\alpha$ the ARCH parameter, $u_{k,t}$ the conditional score with respect to $\eta_{k,t}$ and $\beta$ the
leverage parameter. 
\\
We suppose that the time series in $c=1$ are generated by different D.G.P. (\ref{eq:17}) with respect to those in $c=2$. In particular, we first suppose that the time series in $c=1$ are generated by a D.G.P. with leverage (so $\beta\neq0$) with the same parameters of those in $c=2$ that, instead, is supposed to be generated by a process without leverage. The Fig. \ref{fig:sim1} shows the difference of four time series simulated by the process (\ref{eq:17}) with exactly the same parameters. However, while for two we suppose the leverage ($\beta\neq0$), for the other two we assume no leverage ($\beta=0$).
\\
As evident, the time series generated with the process augmented for the leverage are characterized by similar fluctuations of those without leverage but with shocks (e.g. stock returns immediately after a financial crisis) that dramatically changes the scale of the figures (see Fig. \ref{fig:sim1}).

In a second scenario, instead, we suppose that the time series in the two groups are generated by the D.G.P. in (\ref{eq:17}) both without leverage but with different parameters.
\\
The proposed DCS-A-FCM algorithm is compared with respect the Autocorrelation based Fuzzy C-Means of \cite{durso2009} that is close in the spirit to our proposal. For a fair comparison, besides we know the true D.G.P. (\ref{eq:17}), we perform the classification with a Dynamic Conditional Score model based on a wrong density specification. Indeed, we suppose a Gaussian distribution of the data while they are Skew-t distributed, so we estimate the conditional mean and the conditional variance with a Gaussian-DCS(1,1) as presented in the Section 4.2.1.
\\
The Tab. \ref{tab:sim1} shows the classification accuracy for the first scenario of the proposed algorithm and its comparison with the A-FCM. The considered time series are of different lengths and the the classification is done according to different lags' selection in computing the dissimilarity measure (\ref{eq:11}). 

\begin{table}[H]
    \centering
   \begin{tabular}{rrrrrrr}
    \hline
\multicolumn{4}{c}{A-FCM} \\
    \hline
    Lags of $d_{ACF}$ & $T=50$ & $T=200$ & $T=500$ \\
\hline
     $L=10$ & 61.20\% & 61.70\% & 60.90\% \\
     $L=25$ & 62.08\% & 59.49\% & 65.54\% \\
     $L=50$ & 55.88\% & 61.42\% & 61.97\% \\
    \hline
\multicolumn{4}{c}{DCS-A-FCM ($r=1$)} \\
  \hline
    Lags of $d_{ACF}^{(1)}$ & $T=50$ & $T=200$ & $T=500$ \\
\hline
     $L=10$ & 68.60\% & 64.20\% & 64.50\% \\
     $L=25$ & 65.62\% & 65.71\% & 67.17\% \\
     $L=50$ & 67.05\% & 70.00\% & 67.08\% \\
    \hline
\multicolumn{4}{c}{DCS-A-FCM ($r=2$)} \\
    \hline
    Lags of $d_{ACF}^{(2)}$ & $T=50$ & $T=200$ & $T=500$ \\
\hline
     $L=10$ & 65.80\% & 65.80\% & 70.50\% \\
     $L=25$ & 72.71\% & 70.51\% & 69.78\% \\
     $L=50$ & 71.37\% & 71.42\% & 72.08\% \\
    \hline
    \end{tabular}
    \caption{Classification rate for $M=300$ simulations. The distance measure $d_{ACF}^{(r)}$ is computed at different lags $L$. Both the groups are generated with the D.G.P. (\ref{eq:17}) replacing $\omega=1e^{-6}$, $\phi=0.6$ and $\alpha=0.1$. However, while in the first group $\beta=0$ in the second group $\beta=1$. The classification rate refers to the percentage of the \emph{average} correct classification over the $M=300$ trials.}
    \label{tab:sim1}
\end{table}

First of all it is evident the superior classification ability of the DCS-based clustering procedures with respect the A-FCM for all the considered time series lengths, even if we specified a wrong density function. In particular, the classification based on the conditional variance provides better results (up to 72\% in the case of long time series) than the classification done according to the conditional mean.
\\
The time series length is an important element to evaluate the robustness of the result for either short and long time series. On one hand, the A-FCM algorithm ensures good performances with long time series ($T=500$) with an accuracy close to 65\% in the best case but it is not able to perform better then our proposal also in this case (both with 67\% and 72\% of correct classification rate). On the other hand, for short time series, the DCS-based procedures show robust results with the classification accuracy about the 70\%, while the A-FCM has an accuracy level a bit higher than 50\%.
\\
Moreover, it is not clear what is the effect of the lag's length $L$. Indeed, in the case of short time series length, for the A-FCM algorithm it seems that increasing $L$ reduces the classification accuracy, while for the DCS-A-FCM with the conditional mean similar are robust to the selection of the lag length. In the end, the DCS-A-FCM with the conditional variance ensures higher classification accuracy with increasing $L$. In the case of long time series, instead, for all the classification algorithms it seems that with increasing lag length $L$ the miss-classification rate decreases, with the highest accuracy obtained for all of them specifying $L=50$.
\\
The Tab. \ref{tab:sim2} shows, instead, the classification accuracy for the second scenario where the time series are generated with the D.G.P. (\ref{eq:17}) without leverage and different parameters. As in the previous case, the time series are simulated for different lengths $T$ and the classification is done according to different lags' selection $L$.  

\begin{table}[H]
    \centering
   \begin{tabular}{rrrrrrr}
    \hline
\multicolumn{4}{c}{A-FCM} \\
    \hline
    Lags of $d_{ACF}$ & $T=50$ & $T=200$ & $T=500$ \\
\hline
    $L=10$ & 60.00\% & 56.40\% & 56.00\% \\
     $L=25$ & 56.70\% & 55.90\% & 54.80\% \\
     $L=50$ & 57.20\% & 57.80\% & 58.70\% \\
    \hline
\multicolumn{4}{c}{DCS-A-FCM ($r=1$)} \\
  \hline
    Lags of $d_{ACF}^{(1)}$ & $T=50$ & $T=200$ & $T=500$ \\
\hline
      $L=10$ & 63.60\% & 65.90\% & 63.20\% \\
     $L=25$ & 64.10\% & 64.00\% & 66.30\% \\
     $L=50$ & 63.60\% & 66.60\% & 65.10\% \\
    \hline
\multicolumn{4}{c}{DCS-A-FCM ($r=2$)} \\
    \hline
    Lags of $d_{ACF}^{(2)}$ & $T=50$ & $T=200$ & $T=500$ \\
\hline
     $L=10$ & 67.90\% & 75.90\% & 80.40\% \\
     $L=25$ & 65.60\% & 77.70\% & 82.50\% \\
     $L=50$ & 62.40\% & 76.30\% & 82.80\% \\
    \hline
    \end{tabular}
    \caption{Classification rate for $M=300$ simulations. The distance measure $d_{ACF}^{(r)}$ is computed at different lags $L$. Both the groups have $\beta=0$ but with different parameters. The first group is generated with the D.G.P. (\ref{eq:17}) replacing $\omega=2e^{-6}$, $\phi=0.4$ and $\alpha=0.2$. The second group is generated by the D.G.P. (\ref{eq:17}) assuming $\omega=1e^{-6}$, $\phi=0.8$ and $\alpha=0.05$. The classification rate refers to the percentage of the \emph{average} correct classification over the $M=300$ trials.}
    \label{tab:sim2}
\end{table}

In this second scenario, less influenced by the presence of outliers, the classification accuracy of the DCS-based clustering algorithms is more evident. The benchmark clustering model, the A-FCM, shows in most of the considered cases an accuracy close to 50\%, while the proposed algorithms reach also the 82\% of correct classification rate.
\\
As in the previous example, clustering according to conditional variance provide a much higher classification accuracy with respect the clustering based on the conditional mean.
\\
More in details, for the case of short time series, the over performances of the proposed DCS-A-FCM procedures are higher up to the 10\% of greater classification accuracy, with percentage always much above the 60\% threshold, that is the highest rate for the benchmark A-FCM. In the case of long time series, the conditional moment-based clustering approaches provide over performances up to 26\% of greater classification accuracy rate.
\\
In the end, in this second scenario the results seems more robust to the choice of the lag length $L$ for all the studied clustering algorithms. However, according to the simulations' results, with long time series a lag length $L=50$
it appears the best choice.

\section{Conclusions}

In this paper, we develop a new model-based fuzzy clustering algorithm able to classify time series according to their conditional higher moments, estimated via a Dynamic Conditional Score. Since the implemented statistical model is based on the score, it exploits the complete density structure of the data rather than just a few moments. The similarity between conditional moments' estimates is measured according to a Euclidian distance of the estimated sample autocorrelations. In other words, we propose to cluster time series according to some features of conditional moments. 
\\
Thanks to the generality of the underlying statistical model, the framework presented here generalizes most of the model-based procedures existing in the literature. In this respect, it is worth to mention that a wide class of GARCH-type processes are just special cases of the Dynamic Conditional Score. By specifying different distribution densities for the considered time series, we can cluster challenging instances like non-Gaussian heteroskedastic time series that also show time-varying higher moments. Moreover, the Dynamic Conditional Score allows us to consider also nonlinear models underlying the clustering procedure. The Beta-t-EGARCH is a clear example of a possible nonlinear and non-Gaussian model specification.
\\
Interestingly, we adopt a fuzzy clustering perspective. In doing so, we admit that each time series can be in more than one cluster with a certain level of probability. Indeed, the fuzzy approach implicitly indicates the presence of a second-best cluster -- sometimes, almost as good as the first best; this is a property that is missing in the traditional clustering methods. Moreover, in the real world, the identification of a clear boundary between clusters is not an easy task, so a fuzzy approach appears more attractive than a deterministic one.
\\
We present applications to real financial data based on stock returns. In this respect, we show that it is possible to obtain a number of reference conditional moments for clustering, which depends on the number of time-varying parameters in the specified Dynamic Conditional Score. In the case of a Gaussian-DCS model, we get $R=2$ time-varying moments (mean and variance) with two clusters. In the case of a t-DCS model, instead, we have $R=3$ time-varying parameters and the same number of clusters. Trough the DCS-based fuzzy clustering we could obtain also thinner clusters with time series that share a similar conditional distribution if those series are placed in the same cluster $R$ times.
\\
Future researches could be devoted to the possible applications of the proposed clustering procedure in other real-world frameworks. An interesting proposal might be the application of the presented clustering procedure to the construction of portfolios of stocks when there is a huge amount of available assets. Nevertheless, the clustering procedure we propose is general, and it could also be applied in other disciplines such as engineering, like the relevant case of signal processes clustering.


\begin{thebibliography}{100}
	
	\bibitem{afridi2020}
	M.~K. Afridi, N.~Azam, and J.~Yao.
	\newblock Variance based three-way clustering approaches for handling
	overlapping clustering.
	\newblock {\em International Journal of Approximate Reasoning}, 118:47--63,
	2020.
	
	\bibitem{aghabozorgi2015}
	S.~Aghabozorgi, A.~S. Shirkhorshidi, and T.~Y. Wah.
	\newblock Time-series clustering--a decade review.
	\newblock {\em Information Systems}, 53:16--38, 2015.
	
	\bibitem{alonso2006}
	A.~M. Alonso and E.~A. Maharaj.
	\newblock Comparison of time series using subsampling.
	\newblock {\em Computational statistics \& data analysis}, 50(10):2589--2599,
	2006.
	
	\bibitem{ardia2019}
	D.~Ardia, K.~Boudt, and L.~Catania.
	\newblock Generalized autoregressive score models in r: The gas package.
	\newblock {\em Journal of Statistical Software}, 88(6):1--28, 2019.
	
	\bibitem{ayala2017}
	A.~Ayala, S.~Blazsek, and {\'A}.~Escribano.
	\newblock Dynamic conditional score models with time-varying location, scale
	and shape parameters.
	\newblock 2017.
	
	\bibitem{azzalini2003}
	A.~Azzalini and A.~Capitanio.
	\newblock Distributions generated by perturbation of symmetry with emphasis on
	a multivariate skew t-distribution.
	\newblock {\em Journal of the Royal Statistical Society: Series B (Statistical
		Methodology)}, 65(2):367--389, 2003.
	
	\bibitem{benjamin2003}
	M.~A. Benjamin, R.~A. Rigby, and D.~M. Stasinopoulos.
	\newblock Generalized autoregressive moving average models.
	\newblock {\em Journal of the American Statistical association},
	98(461):214--223, 2003.
	
	\bibitem{bezdek2013}
	J.~C. Bezdek.
	\newblock {\em Pattern recognition with fuzzy objective function algorithms}.
	\newblock Springer Science \& Business Media, 2013.
	
	\bibitem{bollerslev1986}
	T.~Bollerslev.
	\newblock Generalized autoregressive conditional heteroskedasticity.
	\newblock {\em Journal of econometrics}, 31(3):307--327, 1986.
	
	\bibitem{bollerslev1987}
	T.~Bollerslev.
	\newblock A conditionally heteroskedastic time series model for speculative
	prices and rates of return.
	\newblock {\em The review of economics and statistics}, pages 542--547, 1987.
	
	\bibitem{caiado2006}
	J.~Caiado, N.~Crato, and D.~Pe{\~n}a.
	\newblock A periodogram-based metric for time series classification.
	\newblock {\em Computational Statistics \& Data Analysis}, 50(10):2668--2684,
	2006.
	
	\bibitem{caiado2009}
	J.~Caiado, N.~Crato, and D.~Pe{\~n}a.
	\newblock Comparison of times series with unequal length in the frequency
	domain.
	\newblock {\em Communications in Statistics—Simulation and
		Computation{\textregistered}}, 38(3):527--540, 2009.
	
	\bibitem{campello2006}
	R.~J. Campello and E.~R. Hruschka.
	\newblock A fuzzy extension of the silhouette width criterion for cluster
	analysis.
	\newblock {\em Fuzzy Sets and Systems}, 157(21):2858--2875, 2006.
	
	\bibitem{cerqueti2020}
	R.~Cerqueti, M.~Giacalone, and R.~Mattera.
	\newblock Skewed non-gaussian garch models for cryptocurrencies volatility
	modelling.
	\newblock {\em Information Sciences}, 2020.
	
	\bibitem{cerqueti2019}
	R.~Cerqueti, M.~Giacalone, and D.~Panarello.
	\newblock A generalized error distribution copula-based method for portfolios
	risk assessment.
	\newblock {\em Physica A: Statistical Mechanics and its Applications},
	524:687--695, 2019.
	
	\bibitem{cont2001}
	R.~Cont.
	\newblock Empirical properties of asset returns: stylized facts and statistical
	issues.
	\newblock 2001.
	
	\bibitem{coppi2010}
	R.~Coppi, P.~D’Urso, and P.~Giordani.
	\newblock A fuzzy clustering model for multivariate spatial time series.
	\newblock {\em Journal of Classification}, 27(1):54--88, 2010.
	
	\bibitem{corduas2008}
	M.~Corduas and D.~Piccolo.
	\newblock Time series clustering and classification by the autoregressive
	metric.
	\newblock {\em Computational statistics \& data analysis}, 52(4):1860--1872,
	2008.
	
	\bibitem{creal2013}
	D.~Creal, S.~J. Koopman, and A.~Lucas.
	\newblock Generalized autoregressive score models with applications.
	\newblock {\em Journal of Applied Econometrics}, 28(5):777--795, 2013.
	
	\bibitem{durso2000}
	P.~D’Urso.
	\newblock Dissimilarity measures for time trajectories.
	\newblock {\em Statistical Methods \& Applications}, 1(9):53--83, 2000.
	
	\bibitem{durso2013}
	P.~D’Urso, C.~Cappelli, D.~Di~Lallo, and R.~Massari.
	\newblock Clustering of financial time series.
	\newblock {\em Physica A: Statistical Mechanics and its Applications},
	392(9):2114--2129, 2013.
	
	\bibitem{durso2009}
	P.~D’Urso and E.~A. Maharaj.
	\newblock Autocorrelation-based fuzzy clustering of time series.
	\newblock {\em Fuzzy Sets and Systems}, 160(24):3565--3589, 2009.
	
	\bibitem{deluca2011}
	G.~De~Luca and P.~Zuccolotto.
	\newblock A tail dependence-based dissimilarity measure for financial time
	series clustering.
	\newblock {\em Advances in Data Analysis and Classification}, 5(4):323--340,
	2011.
	
	\bibitem{diciccio2011}
	T.~J. Di~Ciccio and A.~C. Monti.
	\newblock Inferential aspects of the skew t-distribution.
	\newblock {\em Quaderni di Statistica}, 13:1--21, 2011.
	
	\bibitem{diaz2010}
	S.~P. D{\'\i}az and J.~A. Vilar.
	\newblock Comparing several parametric and nonparametric approaches to time
	series clustering: a simulation study.
	\newblock {\em Journal of classification}, 27(3):333--362, 2010.
	
	\bibitem{durso2004}
	P.~D'urso.
	\newblock Fuzzy c-means clustering models for multivariate time-varying data:
	different approaches.
	\newblock {\em International Journal of Uncertainty, Fuzziness and
		Knowledge-Based Systems}, 12(03):287--326, 2004.
	
	\bibitem{durso2016}
	P.~D'Urso, L.~De~Giovanni, and R.~Massari.
	\newblock Garch-based robust clustering of time series.
	\newblock {\em Fuzzy Sets and Systems}, 305:1--28, 2016.
	
	\bibitem{durso2018}
	P.~D'Urso, L.~De~Giovanni, and R.~Massari.
	\newblock Robust fuzzy clustering of multivariate time trajectories.
	\newblock {\em International Journal of Approximate Reasoning}, 99:12--38,
	2018.
	
	\bibitem{durso2012}
	P.~D'Urso and E.~A. Maharaj.
	\newblock Wavelets-based clustering of multivariate time series.
	\newblock {\em Fuzzy Sets and Systems}, 193:33--61, 2012.
	
	\bibitem{harvey2014}
	A.~Harvey and G.~Sucarrat.
	\newblock Egarch models with fat tails, skewness and leverage.
	\newblock {\em Computational Statistics \& Data Analysis}, 76:320--338, 2014.
	
	\bibitem{harvey2013}
	A.~C. Harvey.
	\newblock {\em Dynamic models for volatility and heavy tails: with applications
		to financial and economic time series}, volume~52.
	\newblock Cambridge University Press, 2013.
	
	\bibitem{harvey2008}
	A.~C. Harvey and T.~Chakravarty.
	\newblock Beta-t-(e) garch.
	\newblock 2008.
	
	\bibitem{harvey1999}
	C.~R. Harvey and A.~Siddique.
	\newblock Autoregressive conditional skewness.
	\newblock {\em Journal of financial and quantitative analysis}, pages 465--487,
	1999.
	
	\bibitem{huynh2013}
	V.-N. Huynh and V.~Kreinovich.
	\newblock Uncertainty in financial econometrics: Editorial.
	\newblock {\em International Journal of Approximate Reasoning}, 2013.
	
	\bibitem{iorio2018}
	C.~Iorio, G.~Frasso, A.~D’Ambrosio, and R.~Siciliano.
	\newblock A p-spline based clustering approach for portfolio selection.
	\newblock {\em Expert Systems with Applications}, 95:88--103, 2018.
	
	\bibitem{jarque1987}
	C.~M. Jarque and A.~K. Bera.
	\newblock A test for normality of observations and regression residuals.
	\newblock {\em International Statistical Review/Revue Internationale de
		Statistique}, pages 163--172, 1987.
	
	\bibitem{kovsmelj1990}
	K.~Ko{\v{s}}melj and V.~Batagelj.
	\newblock Cross-sectional approach for clustering time varying data.
	\newblock {\em Journal of Classification}, 7(1):99--109, 1990.
	
	\bibitem{krishnapuram2001}
	R.~Krishnapuram, A.~Joshi, O.~Nasraoui, and L.~Yi.
	\newblock Low-complexity fuzzy relational clustering algorithms for web mining.
	\newblock {\em IEEE transactions on Fuzzy Systems}, 9(4):595--607, 2001.
	
	\bibitem{leon2005}
	{\'A}.~Le{\'o}n, G.~Rubio, and G.~Serna.
	\newblock Autoregresive conditional volatility, skewness and kurtosis.
	\newblock {\em The Quarterly Review of Economics and Finance},
	45(4-5):599--618, 2005.
	
	\bibitem{liao2005}
	T.~W. Liao.
	\newblock Clustering of time series data—a survey.
	\newblock {\em Pattern recognition}, 38(11):1857--1874, 2005.
	
	\bibitem{maharaj2010b}
	E.~A. Maharaj and P.~D’Urso.
	\newblock A coherence-based approach for the pattern recognition of time
	series.
	\newblock {\em Physica A: Statistical mechanics and its Applications},
	389(17):3516--3537, 2010.
	
	\bibitem{maharaj2011}
	E.~A. Maharaj and P.~D’Urso.
	\newblock Fuzzy clustering of time series in the frequency domain.
	\newblock {\em Information Sciences}, 181(7):1187--1211, 2011.
	
	\bibitem{maharaj2010}
	E.~A. Maharaj, P.~D’Urso, and D.~U. Galagedera.
	\newblock Wavelet-based fuzzy clustering of time series.
	\newblock {\em Journal of classification}, 27(2):231--275, 2010.
	
	\bibitem{maharaj2019}
	E.~A. Maharaj, P.~D'Urso, and J.~Caiado.
	\newblock {\em Time series clustering and classification}.
	\newblock CRC Press, 2019.
	
	\bibitem{nelson1991}
	D.~B. Nelson.
	\newblock Conditional heteroskedasticity in asset returns: A new approach.
	\newblock {\em Econometrica: Journal of the Econometric Society}, pages
	347--370, 1991.
	
	\bibitem{otranto2008}
	E.~Otranto.
	\newblock Clustering heteroskedastic time series by model-based procedures.
	\newblock {\em Computational Statistics \& Data Analysis}, 52(10):4685--4698,
	2008.
	
	\bibitem{otranto2010}
	E.~Otranto.
	\newblock Identifying financial time series with similar dynamic conditional
	correlation.
	\newblock {\em Computational Statistics \& Data Analysis}, 54(1):1--15, 2010.
	
	\bibitem{piccolo1990}
	D.~Piccolo.
	\newblock A distance measure for classifying arima models.
	\newblock {\em Journal of Time Series Analysis}, 11(2):153--164, 1990.
	
	\bibitem{russell2005}
	J.~R. Russell and R.~F. Engle.
	\newblock A discrete-state continuous-time model of financial transactions
	prices and times: The autoregressive conditional multinomial--autoregressive
	conditional duration model.
	\newblock {\em Journal of Business \& Economic Statistics}, 23(2):166--180,
	2005.
	
	\bibitem{thadewald2007}
	T.~Thadewald and H.~B{\"u}ning.
	\newblock Jarque--bera test and its competitors for testing normality--a power
	comparison.
	\newblock {\em Journal of applied statistics}, 34(1):87--105, 2007.
	
	\bibitem{theodossiou2015a}
	P.~Theodossiou.
	\newblock Skewed generalized error distribution of financial assets and option
	pricing.
	\newblock {\em Multinational Finance Journal}, 19(4):223--266, 2015.
	
	\bibitem{urzua1996}
	C.~M. Urz{\'u}a.
	\newblock On the correct use of omnibus tests for normality.
	\newblock {\em Economics Letters}, 53(3):247--251, 1996.
	
	\bibitem{wisniewska2017}
	M.~Wi{\'s}niewska and A.~Wy{\l}oma{\'n}ska.
	\newblock Garch process with ged distribution.
	\newblock In {\em Cyclostationarity: Theory and Methods III}, pages 83--103.
	Springer, 2017.
	
\end{thebibliography}

\end{document}